\documentclass{emulateapj}
\newcommand{\eg}{{\it e.g.}}    
\newcommand{\ie}{{\it i.e.}}    
\newcommand{\etal}{et al.}      
\newcommand{\inv}{$^{-1}$}
\newcommand{\kms}{km~s\inv}
\newcommand{\vi}{$V_{606}-i_{775}$}
\newcommand{\HST}{\emph{HST}}
\newcommand{\sextractor}{\textsc{SExtractor}}
\slugcomment{Submitted to The Astronomical Journal}
\shorttitle{Treasury Redshift Survey of the GOODS-North Field}
\shortauthors{Team Keck}
\begin{document}
\title{The Team Keck Treasury Redshift Survey of the GOODS-North
  Field\altaffilmark{1}}
\author{Gregory~D.~Wirth\altaffilmark{2},
  Christopher~N.~A.~Willmer\altaffilmark{3,4}, 
  Paola~Amico\altaffilmark{2}, 
  Frederic~H.~Chaffee\altaffilmark{2}, 
  Robert~W.~Goodrich\altaffilmark{2}, 
  Shui~Kwok\altaffilmark{2}, 
  James~E.~Lyke\altaffilmark{2}, 
  Jeff~A.~Mader\altaffilmark{2}, 
  Hien~D.~Tran\altaffilmark{2}, 
  Amy~J.~Barger\altaffilmark{5,6}, 
  Lennox~L.~Cowie\altaffilmark{5}, 
  Peter~Capak\altaffilmark{5}, 
  Alison~L.~Coil\altaffilmark{7}, 
  Michael~C.~Cooper\altaffilmark{7}, 
  Al~Conrad\altaffilmark{2}, 
  Marc~Davis\altaffilmark{7}, 
  S.~M.~Faber\altaffilmark{3}, 
  Esther~M.~Hu\altaffilmark{5}, 
  David~C.~Koo\altaffilmark{3}, 
  David~Le~Mignant\altaffilmark{2}, 
  Jeffrey~A.~Newman\altaffilmark{7},
  Antoinette~Songaila\altaffilmark{5}} 
\altaffiltext{1}{Based in part on data obtained at the W. M. Keck
  Observatory, which is operated as a scientific partnership among the
  California Institute of Technology, the University of California,
  and NASA, and was made possible by the generous financial support of
  the W. M. Keck Foundation.}
\altaffiltext{2}{W. M. Keck Observatory, Kamuela, HI 96743; \texttt{
    zee@keck.hawaii.edu}.}
\altaffiltext{3}{University of California Observatories/Lick Observatory,
  Department of Astronomy \& Astrophysics, University of California,
  Santa Cruz, CA 95064.}
\altaffiltext{4}{On leave from Observat\'orio Nacional, Brazil.}
\altaffiltext{5}{Institute for Astronomy, University of Hawaii, 2680
  Woodlawn Drive, Honolulu, HI 96822.}
\altaffiltext{6}{Department of Astronomy, University of
  Wisconsin-Madison, 475 North Charter Street, Madison, WI 53706.}
\altaffiltext{7}{Department of Astronomy, University of California,
  Berkeley, CA 94720.}

\begin{abstract}
  We report the results of an extensive imaging and spectroscopic
  survey in the GOODS-North field completed using DEIMOS on the Keck II
  telescope.  Observations of 2018 targets in a magnitude-limited
  sample of 2911 objects to $R_{AB}=24.4$ yield secure redshifts for a
  sample of 1440 galaxies and AGN plus 96 stars.  In addition to
  redshifts and associated quality assessments, our catalog also
  includes photometric and astrometric measurements for all targets
  detected in our $R$-band imaging survey of the GOODS-North region.  We
  investigate various sources of incompleteness and find the redshift
  catalog to be 53\% complete at its limiting magnitude.  The median
  redshift of $z=0.65$ is lower than in similar deep surveys
  because we did not select against low-redshift targets.
  
  Comparison with other redshift surveys in the same field, including
  a complementary Hawaii-led DEIMOS survey, establishes that our
  velocity uncertainties are as low as $\sigma\approx40$~\kms\ for red
  galaxies and that our redshift confidence assessments are accurate.
  The distributions of rest-frame magnitudes and colors among the
  sample agree well with model predictions out to and beyond $z=1$.
  We will release all survey data, including extracted 1-D and
  sky-subtracted 2-D spectra, thus providing a sizable and homogeneous
  database for the GOODS-North field which will enable studies of large
  scale structure, spectral indices, internal galaxy kinematics, and
  the predictive capabilities of photometric redshifts.
\end{abstract}


\section{Introduction}\label{sxn-intro}

In 1995, the Hubble Deep Field (HDF) became the most well-studied
extragalactic area in all of observational astronomy.  Extremely deep
optical exposures taken with the \emph{Hubble Space Telescope (HST)}
were eventually joined by additional imaging and spectroscopic
observations from many observatories covering all wavelengths (cf.
\cite{fer00} and references therein).  This combined dataset --- which
was ultimately expanded to include a complementary field in the
southern sky --- painted an extraordinarily detailed picture of a
``blank'' region of sky which turned out to be teeming with galaxies,
providing astronomers with an exquisite window into the high-redshift
universe.  The result was a series of important advances in the
properties of distant galaxies and large-scale structure in the
universe.  Many of these key discoveries depended critically upon
knowledge of galaxy redshifts, the vast majority of which were
obtained by a variety of observers using the LRIS spectrograph
\citep{oke95} on the twin 10~m telescopes of the W. M. Keck
Observatory
\citep{coh96,cow96,ste96,low97,phi97,mou97,coh00,coh01,daw01}.

Nearly ten years later, attention has returned to the northern HDF
field (HDF-N) via an ambitious plan to use a new generation of ground-
and space-based instruments for probing the distant universe in even
greater detail.  The Great Observatories Origins Deep Survey (GOODS;
\cite{dic01}) project involves a coordinated effort by three of NASA's
Great Observatories to obtain ultra-deep images of two selected fields
in the optical (via the \HST\ Advanced Camera for Surveys), infrared
(with the \emph{Spitzer Space Telescope}), and X-ray (using
\emph{Chandra}) wavelength regimes.  These two fields include a
northern target area coincident with (but significantly larger than)
the HDF-N plus a similarly-sized southern field coincident with the
Chandra Deep Field-South.  Again, intensive ground-based followup
observations in all wavelength regimes are underway to exploit this
high-resolution space-based imaging.  High-quality spectra of the many
galaxies in the GOODS fields will, as before, be essential to
understand both large-scale structure and galaxy evolution within
these regions.

Keck remains the most powerful northern-hemisphere observatory for
completing deep surveys of faint galaxies.  Its efficient new DEIMOS
spectrograph \citep{fab03} on the Keck~II telescope was designed
specifically for acquiring multislit spectroscopy of over 100 faint
galaxies at once, and with its $16\arcmin$ field of view is ideally
suited to survey the $10\arcmin\times16\arcmin$ GOODS-North (hereafter
GOODS-N) field.  Presented with this unique opportunity to make an
important contribution to the community, the WMKO Director and
scientific staff elected to complete a deep DEIMOS survey of this
region and committed to releasing the resulting data to the community
as rapidly as possible.  We report herein the results of this ``Team
Keck'' Treasury Redshift Survey (hereafter, TKRS) in the GOODS-N
field.  Our survey is complemented by the Hawaii-led Keck/DEIMOS
survey in GOODS-N described in the accompanying report \citep{cow04},
and also by a Gemini-North survey completed with GMOS \citep{dic04}.

In \S\ref{sxn-photo} of this report we review the construction of
the photometric catalog of the GOODS-N field on which we based our
slitmask designs.  Section~\ref{sxn-spec} discusses the acquisition
and reduction of the spectra and the determination of redshifts.
Section~\ref{sxn-catalog} concerns the redshift catalog, including the
effects of target characteristics on the sample's completeness, the
comparison of our survey with others in the same area, the accuracy of
our redshifts and quality assessments, and the accuracy of photometric
redshifts in this field.  Section~\ref{sxn-analysis} presents the
redshift and rest-frame color distribution for our targets, and
\S\ref{sxn-summary} concludes the paper.

\section{Photometric Catalog}\label{sxn-photo}

To ensure timely access to a catalog of astrometric positions and
approximate magnitudes for completing the spectroscopy, we generated
our own independent catalog of sources in the GOODS-N field.  We
acquired a series of $300$~s $R$-band exposures using DEIMOS in
imaging mode on Keck II during the night of 2002 Dec 30 (UT), and
complemented these with additional images obtained on 2003 Feb 02 (UT)
to complete coverage of the full ACS imaging region.  The exposures
were offset and dithered to compensate for the gaps in the DEIMOS CCD
mosaic.

All images were bias subtracted and gain corrected, then flatfielded
with sky flats generated from the set of 14 images acquired in this
field.  We employed an automated procedure to detect cosmic rays on
each image and ignored the affected pixels in all subsequent
processing.  The mean sky level on each image was measured via an
iterative sigma clipping algorithm and subtracted to eliminate sky
brightness variations from the data set.  Next, we used DEIMOS images
of an astrometric field to derive a distortion correction map which we
applied to the images of the GOODS-N field.  This allowed us to shift
and co-add the science images, yielding a mosaic image that covered
the entire $10\arcmin\times16\arcmin$ first-epoch ACS imaging field.
For this survey, we only considered the portion of this mosaic image
coincident with the first epoch \HST\ GOODS-N observations.  We
obtained a first-order plate solution for this mosaic using several
stars measured in the USNO astrometric catalog \citep{mon03}.

To generate our target catalog we used the \sextractor\ package
\citep{ber96} to detect positive fluctuations at least 20~px in size
with a threshold of 1.0 times the RMS sky noise.  Because of
vignetting near the edges of the DEIMOS imaging field of view, the
final mosaic image contains a few regions with spurious detections.
These bogus objects are generally very faint, and most were excluded
from the final catalogue because they lie below the faint magnitude
cutoff of the sample.

Although the USNO catalog provided an effective first-order
astrometric solution for the GOODS-N mosaic image, the small number of
sources available in this field limited the accuracy of the solution.
The resultant RMS positional uncertainty of approximately $0\farcs5$
is a significant fraction of the $1\arcsec$ slit width for
spectroscopy, and hence a more accurate solution was needed to produce
acceptable slitmask designs.  In order to improve the object
positions, we used a catalogue of objects in this region kindly
provided by the SDSS team \citep{yor00} which produced a list of 130
stars common to both catalogs.  This permitted us to refine the
astrometric solution and obtain a final RMS residual of
$\sim0\farcs13$, more than adequate for our needs.  Comparing our
derived astrometric coordinates to those obtained independently by
\cite{cap03}, we find no measureable offset in ($\alpha,\delta$) and a
dispersion of $\sigma=0\farcs40$, indicating excellent agreement.  Our
coordinates agree less well in absolute zero point with those
\cite{gia04} obtained from \HST\ imaging, against which the dispersion
is $0\farcs23$ but the offsets are $\Delta\alpha=0\farcs00$ and
$\Delta\delta=0\farcs38$; this suggests a slight error in absolute
astrometric positions for the ACS catalog.  For several saturated
stars which we wanted to use for rough slitmask alignment purposes, we
derived estimated celestial coordinates positions visually by
estimating the ($\alpha,\delta$) coordinates for the location at the
convergence points of the diffraction spikes.

\begin{figure}[!bp]
  \includegraphics[angle=270,width=0.475\textwidth]{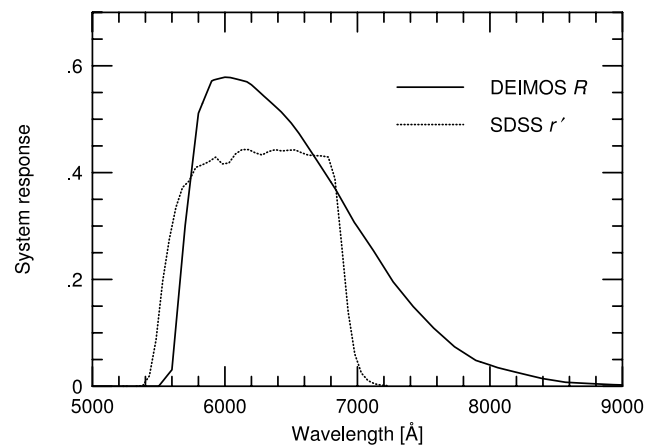}
  \caption{\label{fig-filters}
    System (telescope, filter, and detector) response curves of $R$
    band filter used in the DEIMOS imaging (\emph{solid line}) and the
    SDSS $r'$ filter (\emph{dotted line}).  The $r'$ filter is part of
    the system devised to calibrate the SDSS survey \protect\citep{smi02}, and
    is shown here to illustrate the differences in bandpass of the
    filters.  Note that in the actual calibration of the TKRS
    catalogue, the $r$-band data from the SDSS DR1 release were used.}
\end{figure}

The unsaturated stars also served as photometric calibrators. The
magnitudes we used were \texttt{MAG\_BEST} calculated by \sextractor,
which are designed to estimate the total light.  Since only $r'$
magnitudes were available from the SDSS database, we were unable to
include any color terms in our photometric calibration. For the 130
stars shared between the TKRS and SDSS catalogs, the RMS difference is
$\Delta R = 0.36$~mag. The final magnitudes are in the AB system
\citep{oke74}, roughly equivalent to the SDSS $r'$.
Figure~\ref{fig-filters} compares the SDSS $r'$ and DEIMOS $R$
instrumental response functions.

\begin{figure}[tbp]
  \includegraphics[angle=270,width=0.475\textwidth]{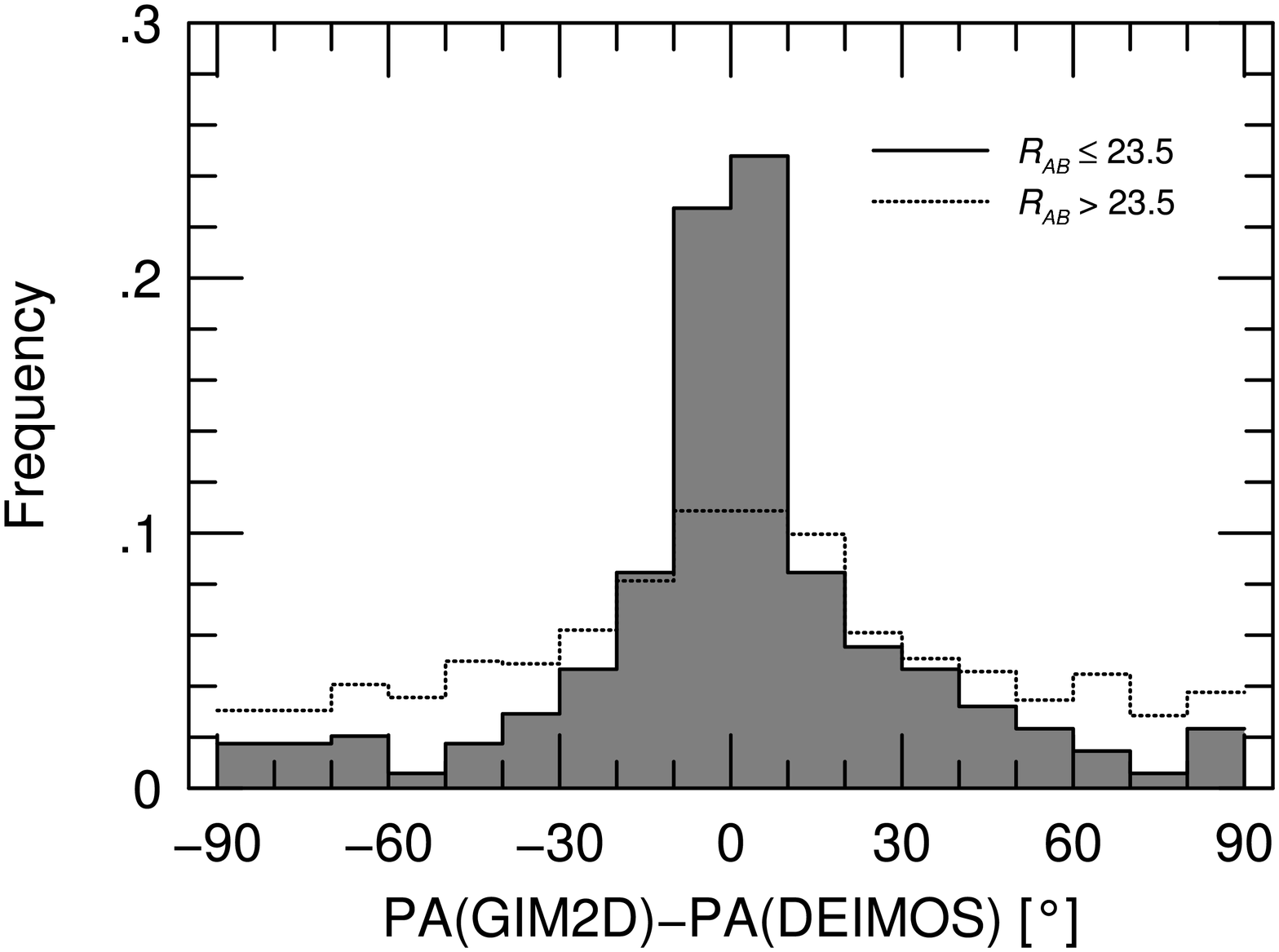}
  \caption{\label{fig-posang}
    Distribution of the difference in position angle measured from
    ground-based DEIMOS $R$-band images vs.\ \HST\ WFPC2 $I$-band
    images in the HDF and Flanking Fields \protect\citep{mar98,sim02}, binned
    in $10\degr$ intervals.  The solid histogram represents brighter
    galaxies with $R_{AB}\leq23.5$, while the solid line represents
    the position angle differences for galaxies fainter than 23.5.
    Both bright and faint galaxies show a peak at position angle
    differences at $0\degr$, but whereas this peak is well defined for
    the brighter galaxies, for fainter ones the smoother distribution
    indicates a poor correlation.  }
\end{figure}

\sextractor\ also provides for each object an estimate of its
orientation as represented by the major axis position angle.  This
information can be used to orient spectroscopic slits to the major
axis of the galaxy and thus obtain both improved $S/N$ spectra and
more accurate rotation curves.  We compared the position angles
derived from the DEIMOS mosaic image with independent
measurements\footnote{Data available at
  \url{http://archive.deep.ucolick.org}} obtained from \HST\ images by
the UCSC DEEP1 team in the Hubble Deep Field and Flanking Fields using
two-dimensional two-parameter fits to the images \citep{mar98,sim02}.
A histogram comparing the difference in position angles using
$10\degr$ bins is shown in Fig.~\ref{fig-posang}. The solid histogram
shows the distribution measured for galaxies with $R\leq23.5$, while
the dotted line represents the fainter galaxies.  Both distributions
have been normalized by the total number of galaxies in each
subsample. Although both distributions are peaked at values close to
$0\degr$, indicating that the two datasets generally agree, the peak
is far more pronounced for the brighter galaxy sample.  We conclude
that the position angles measured by \sextractor\ are adequate for the
brighter galaxies.

\section{Spectroscopy}\label{sxn-spec}
\subsection{Slitmask design}

The selection of objects for spectroscopic observations followed
roughly the same guidelines as used by the DEEP2 Redshift Survey
\citep{dav03}, with one major difference: no attempt was made to use
brightness, color, or other information to exclude stars and
lower-redshift galaxies from the sample.  All objects brighter than
the limiting magnitude were potential spectroscopic targets,
regardless of whether their appearance was stellar or galactic. At the
magnitude limit of this survey the number of stars relative to the
number of galaxies is small, so that the adopted strategy avoids
excluding AGN-dominated galaxies.

\begin{figure*}[tbp]
  \includegraphics[angle=270,width=\textwidth]{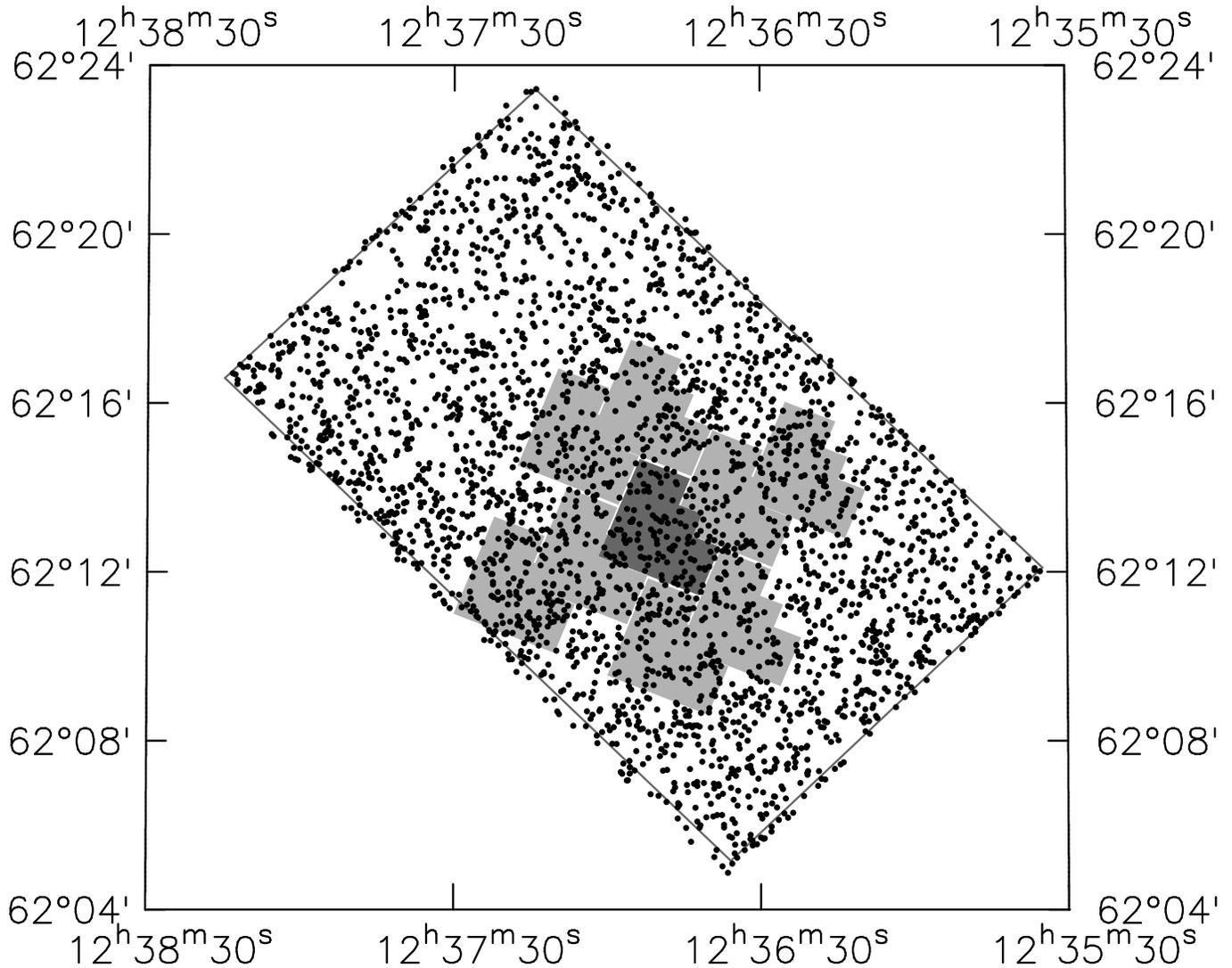}
  \caption{\label{fig-radec-all}
    Projected sky distribution of objects in the DEIMOS spectroscopic
    survey sample.  Shaded areas indicate the Hubble Deep Field
    (\emph{dark grey}) and the flanking fields (\emph{light grey}).
    The outline indicates approximately the first-epoch \HST\ ACS
    GOODS-N imaging survey region.
    }
\end{figure*}

DEIMOS uses slitmasks to record over 100 broadband spectra at once.
The limiting magnitude of the spectroscopic sample, the number of
available nights and uniformity of wavelength coverage were all
considered in designing slitmasks for our survey.  The magnitude
cutoff was set at $R\leq24.4$ in order to provide sufficient source
density per slitmask, yielding a sample of 2911 candidate targets for
inclusion on masks.  To improve consistency of wavelength coverage,
objects were preferentially placed near the center of the $4\arcmin$
available in the spatial direction on each DEIMOS mask.  Automated
slitmask design software developed for the DEEP2 survey then selected
appropriate targets from the catalog to produce a set of 18 slitmasks,
each displaced from the next by $\sim0\farcm33$, to cover the GOODS-N
field.  The resulting distribution in equatorial coordinates of the
2018 objects in the spectroscopic sample is shown in
Fig.~\ref{fig-radec-all}.

All slits were tilted by a minimum amount with respect to the CCD rows
in order to improve the wavelength sampling of the spectra; the slight
offset between adjacent columns in the spectra serves to dither the
night sky lines and thus permit a more accurate reconstruction of the
background spectrum.  For objects exhibiting significant elongation in
the DEIMOS images, slits were typically aligned with the major axis to
improve both $S/N$ and rotational velocity measurements.  For closely
spaced objects, slit position angles were chosen such that the spectra
of both objects were recorded.  However, slits were never tilted by
more than $\pm30\degr$ from the nominal position angle of the
slitmask.

\subsection{Observations}

Spectroscopic observations of the GOODS-N field were completed using
DEIMOS on the Keck~II telescope during early-to-mid 2003 as detailed
in Table~\ref{tab-obslog}.  Each observation used the 600~l~mm\inv\ 
grating blazed at 7500~\AA, with the GG455 order-blocking filter
eliminating all flux below 4550~\AA.  The central wavelength was set
at 7200~\AA, providing nominal spectral coverage of 4600--9800~\AA\ at
a FWHM resolution of $\Gamma\approx3.5$~\AA.  Eleven masks were
observed during time allocated to the WMKO Director, and the remaining
seven were generously completed by the DEEP2 Redshift Survey Team.
Each slitmask was observed for a total on-source integration time of
3600~s, broken up into $3\times1200$~s integrations to allow rejection
of cosmic rays.  No dithering was performed between exposures because
of the use of tilted slits and because the minor fringing pattern
present in DEIMOS images is sufficiently corrected by the use of flat
field images.

\begin{deluxetable*}{ccccrrrr}[tbp]
  \tablewidth{0pc}
  \tablecaption{Slitmask Observation Data\label{tab-obslog}}
  \tablehead{
    \colhead{Mask}         &
    \colhead{Obs. Date}         &
    \colhead{$\alpha$\tablenotemark{a}} &
    \colhead{$\delta$\tablenotemark{a}} &
    \colhead{PA\tablenotemark{b}} &
    \colhead{$N_o$\tablenotemark{c}}    &
    \colhead{$N_z$\tablenotemark{d}} &
    \colhead{$F_z$\tablenotemark{e}} \\
    &(UT)&(J2000)&(J2000)&\colhead{($\degr$)}&&&\colhead{(\%)}}
  \startdata
  1  &  2003 May 02 & 12~36~30.81 & 62~17~01.9 &     44&117 &   72 &  62\\
  2  &  2003 May 02 & 12~36~33.44 & 62~16~33.9 &     44&118 &   91 &  77\\
  3  &  2003 May 02 & 12~36~36.99 & 62~16~20.2 &     44&118 &   94 &  80\\
  4  &  2003 May 03 & 12~36~39.63 & 62~15~52.2 &     44&111 &   97 &  87\\
  5  &  2003 May 03 & 12~36~43.18 & 62~15~38.5 &     44&112 &   99 &  88\\
  6  &  2003 May 03 & 12~36~45.81 & 62~15~10.5 &     44&103 &   94 &  91\\
  7  &  2003 May 04 & 12~36~49.37 & 62~14~56.9 &     44&103 &   69 &  67\\
  8  &  2003 May 05 & 12~36~52.00 & 62~14~28.8 &     44&105 &   80 &  76\\
  9  &  2003 May 06 & 12~36~55.56 & 62~14~15.2 &     44&110 &   89 &  81\\
  10 &  2003 Mar 26 & 12~36~58.19 & 62~13~47.1 &     44&105 &   62 &  59\\
  11 &  2003 Mar 26 & 12~37~01.75 & 62~13~33.5 &     44&114 &   88 &  77\\
  12 &  2003 Mar 26 & 12~37~04.38 & 62~13~05.5 &     44&112 &   93 &  83\\
  13 &  2003 Mar 26 & 12~37~07.93 & 62~12~51.8 &     44&102 &   73 &  72\\
  14 &  2003 Mar 26 & 12~37~10.57 & 62~12~23.8 &     44&118 &   75 &  64\\
  15 &  2003 May 28 & 12~37~14.12 & 62~12~10.1 &     44&120 &  104 &  87\\
  16 &  2003 May 28 & 12~37~16.75 & 62~11~42.1 &     44&125 &   98 &  78\\
  17 &  2003 May 29 & 12~37~20.31 & 62~11~28.5 &     44&116 &   86 &  74\\
  18 &  2003 May 29 & 12~37~19.86 & 62~11~21.3 & $-136$&109 &   72 &  66\\
  \enddata
\tablenotetext{a}{Celestial coordinates of the nominal slitmask center.}
\tablenotetext{b}{Position angle of the slitmask; individual slit
  orientations vary.}
\tablenotetext{c}{Number of objects per mask; note that a slit may
  contain multiple objects.}
\tablenotetext{d}{Number of secure redshifts ($Q=-1$, 3, or 4) measured per mask.}
\tablenotetext{e}{Percentage of objects per mask yielding secure redshifts.}
\end{deluxetable*}

To enable wavelength calibration and the removal of instrumental
signatures, calibration data consisting of three internal quartz flats
and a single arc lamp spectrum (Kr, Ar, Ne, Xe) were obtained for each
mask in the daytime; the DEIMOS flexure compensation system ensured
that these calibration images matched the science frames to better
than $\pm0.25$~px.  The transparency and seeing conditions on Mauna
Kea varied from fair to excellent during the observations, affecting
the resultant data quality.

\subsection{Data reduction}

All spectra were extracted using the fully-automated pipeline
developed by the DEEP2 Redshift Survey Team \citep{new04} and
generously made available to us.  Each spectral exposure was divided
by its corresponding flatfield image to remove small amounts of CCD
fringing.  The three exposures for each mask were then combined to
yield an image cleaned of cosmic rays.  For each slitlet, a 2-D
sky-subtracted spectrum is obtained by modeling the night-sky with a
B-spline and subtracting it from the combined 2-D spectrum.
Wavelength calibration was achieved by fitting to the arc lamp
emission lines.  The sky-subtracted spectrum of the target was
produced by summing over the rows containing the flux from the object
as predicted by the tabulated position relative to the slit endpoints.

\subsection{Redshift determination}

The first phase of measuring redshifts involved fully automated
processing of each spectrum using the same algorithm as the SDSS
\citep{str02}, which calculates for each lag position (\ie, redshift) the
best representation of the source spectrum through the use of a set of
template eigenspectra \citep{gla98}.  These templates include several types of
stars, a galaxy absorption spectrum, a galaxy emission-line spectrum,
and an AGN spectrum. The best fit is determined from the minimum
$\chi^2$ in the redshift-lag and eigenspectra parameter space.  The
ten best fits are stored in the 1-D image header.

In the second phase, all spectra were visually inspected by at least
two team members, who would select both a redshift and a corresponding
quality code; objects for which no unambiguous redshift could be
determined were also noted.  Due to low $S/N$ spectra or to poor
matches between the spectra of the templates and some program objects,
the automated redshift determination algorithm sometimes selected an
incorrect redshift (Newman, private communication).  However, all
redshifts reported herein were agreed upon by two reviewers, and can
thus be considered secure.  The adopted redshift quality codes
(hereafter, $Q$; see Table~\ref{tab-zqual2}) are the same as used by
the DEEP2 team \citep{new04}.  We consider the accuracy of these
quality codes in \S\ref{sxn-zquality}.

\begin{deluxetable*}{rp{0.5\textwidth}rrrrr}[btp]
  \tablecaption{TKRS Redshift Quality Distribution\label{tab-zqual2}}
  \tablehead{
    \colhead{$Q$\tablenotemark{a}}                         &
    \colhead{Definition}                         &
    \colhead{$N_{obj}$\tablenotemark{b}}                   &
    \colhead{$F_{obj}$\tablenotemark{c}}              &
    \colhead{$N_{zo}$\tablenotemark{d}}                     &
    \colhead{$N_{agree}$\tablenotemark{e}}     &
    \colhead{$F_{agree}$\tablenotemark{f}}\\
    &&&\colhead{(\%)}&&&\colhead{(\%)}}
  \startdata
  4 &  Very secure redshift ($P>99\%$); at least two
  spectral features identified & 1180 &  58.5 & 154  & 145 & 94.2 \\
  3 &  Secure redshift ($P>95\%$); one strong line and
  another weak feature identified \emph{or} single wide line, typically
  [\ion{O}{2}] $\lambda3727$ blend &  260 &  12.9 &  83  &  76 & 91.6 \\ 
  2 &  Uncertain redshift; signal is present but no unambiguous
  spectral line identified &  268 &  13.3 &  69  &  26 & 37.7 \\
  1 & No redshift; $S/N$ too poor
  & 183 &   9.1 &  28  &   1 &  3.6 \\
  $-1$ & Star & 96 &  4.8 &  50  &  49 & 98.0 \\
  $-2$ & No redshift measured because of instrumental artifacts in
  spectrum & 31 &  1.5 &   6  &   0 &  0.0 \\
  \enddata
\tablenotetext{a}{TKRS redshift quality category.}
\tablenotetext{b}{Number of objects in TKRS catalog for this category.}
\tablenotetext{c}{Fraction (percentage) of objects in TKRS catalog for this category.}
\tablenotetext{d}{Number of TKRS objects in this category which have
  independent redshift measurements for comparison.} 
\tablenotetext{e}{Number of TKRS objects for which the independent
  redshifts agree to within $|\Delta z|\leq0.002$.}
\tablenotetext{f}{Fraction (percentage) of TKRS objects for which the
  independent  
  redshifts agree to within $|\Delta z|\leq0.002$.}
\end{deluxetable*}

The results from the visual inspections were then collated and
discrepant cases noted and reassessed by one individual.  Here,
``discrepant'' means that one observer assigned a $Q$ value of 3 or 4
(corresponding to a ``secure'' redshift) whereas another would select
$Q=2$ or less, or that two observers would assign secure $Q$ values
but discrepant redshifts.  The latter cases resulted when spectra of
two different galaxies overlapped in the same slit.

The distribution of redshift quality codes among the 2018 observed
targets is shown in Table~\ref{tab-zqual2}. The number of secure
redshift measurements ($Q=-1,3,4$) is 1536.  The $Q=-2$ objects are
technically not counted as failures; hence, the overall efficiency is
77\% (1536/1987), which is comparable to the success rate of DEEP2 and
similar redshift surveys (\eg, \cite{cow04}).

\section{Redshift Catalog}\label{sxn-catalog}

\subsection{Data}

The measurements resulting from our survey are presented in
Table~\ref{tab-tkrs}, a portion of which is reproduced here.  This
table appears in its entirety in the electronic version of the Journal
and also on the website devoted to the
survey\footnote{\url{http://www2.keck.hawaii.edu/science/tksurvey}}.
Photometric and spectroscopic data for all targets detected in our
imaging survey appear here, in addition to alternate redshifts from
other surveys when available.  Redshifts for targets in our survey are
only given when they are classified as secure ($Q=-1,3,4$).  Also
available on the survey website is the master table of measurements
which combines the results of this survey with those of the parallel
effort by \cite{cow04}.  Users of these data should note that the
tabulated redshifts for our survey currently lack a heliocentric
correction; an adjustment will be applied in future releases, thus
changing the zero point for the redshifts by no more than a few tens
of \kms.

\subsection{Sample completeness}

In the present section we examine the systematics of the present
sample as a function of location in the DEIMOS field of view, apparent
magnitude, color, and surface brightness.  The number of objects
placed on each mask appears in Table~\ref{tab-obslog}, which also
shows both the number of targets for which secure redshifts were
obtained ($N(Q=-1,3,4)$) and the success fraction per mask.  On average,
each mask contained about 112 objects, of which 76 (68\%) produced
secure redshifts. However, these numbers do not take into account the
$Q=-2$ objects representing failures due to instrumental effects, so
these numbers understate the survey's redshift measuring efficiency.

\begin{figure}[tbp]
  \includegraphics[angle=270,width=0.475\textwidth]{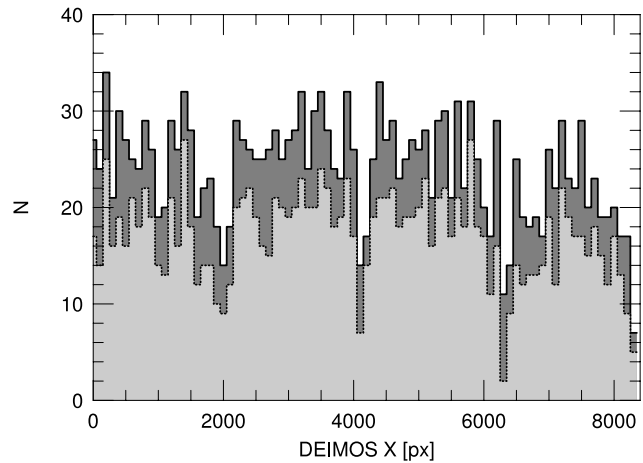}
  \caption{\label{fig-zqual-x}
    Number of targets placed on masks (\emph{dark grey histogram}),
    and number of successful redshift measures (\emph{light grey}) as
    a function of the DEIMOS catalog $X$ coordinate, which is parallel
    to the long axis of the GOODS-N field.  As explained in the text,
    the depressions at $X=2000$, 4000 and 6000 correspond to the gaps
    between CCDs.}
\end{figure}

\begin{figure}[tbp]
  \includegraphics[angle=270,width=0.475\textwidth]{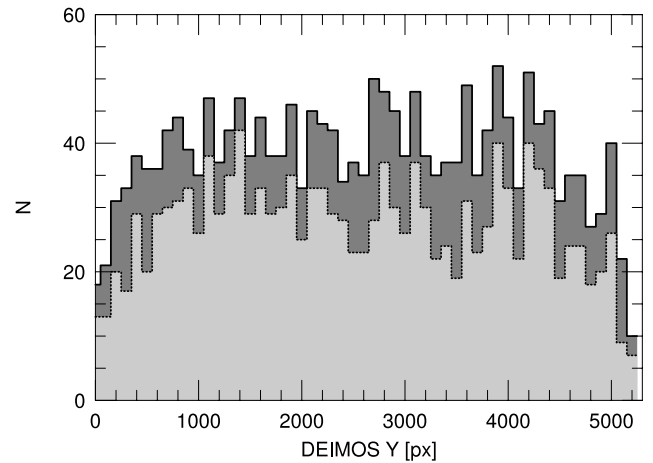}
  \caption{
    Number of targets placed on masks (\emph{dark grey histogram}),
    and number of successful redshift measures (\emph{light grey}), as
    a function of DEIMOS catalog $Y$ coordinate, which is parallel to
    the GOODS-N shorter axis. The fall-off at both edges is caused by
    the fact that galaxies and stars in these positions have a smaller
    probability of being placed on masks, since they are only
    considered as potential mask candidates a smaller number of times.
    \label{fig-zqual-y}
    }
\end{figure}

\paragraph{Spatial effects.} Figures~\ref{fig-zqual-x} and
\ref{fig-zqual-y} depict the distribution of galaxies as a function of
the DEIMOS mosaic $x$ and $y$ coordinates, respectively.  The figures
compare the distribution of all targets (\emph{dark grey histograms})
to those which yielded secure redshifts (\emph{light grey histogram}).
In Fig.~\ref{fig-zqual-x}, the periodic depressions result in part
from a lack of slits placed in the inter-CCD gap regions (a product of
the slitmask design) and in part due to the decreased success rate for
spectra taken acquired too close to the gaps and ruined.  The decrease
in number of objects at the edges of the DEIMOS mosaic $y$ (shorter)
axis is clearly seen in Fig.~\ref{fig-zqual-y}, and this is probably
an effect caused by the slitmask design algorithm. For galaxies and
stars located at the edges there is only one chance of being placed on
masks, in contrast to objects further from the edge.

\begin{figure*}[tbp]
  \includegraphics[angle=270,width=\textwidth]{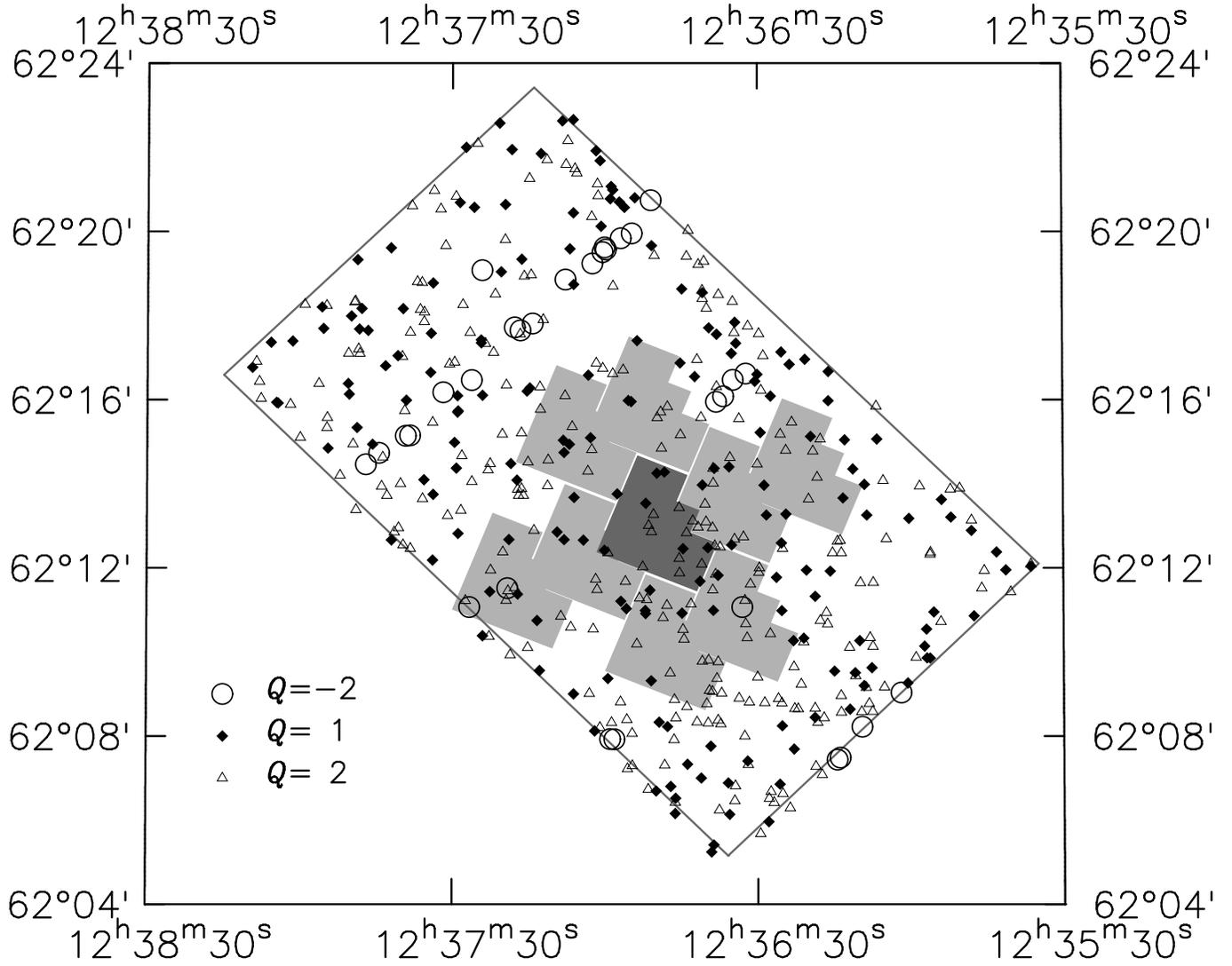}
  \caption{\label{fig-zfailure}
    Projected sky distribution of objects for which no successful
    redshifts were measured.  Plotting symbols distinguish objects for
    which no spectra were obtained because they fell on bad
    columns, gaps between CCDs, or at the edges of masks ($Q=-2$),
    objects for which a spectrum was obtained, but no redshift could
    be deduced ($Q=1$), and objects which may yet produce
    redshifts or for which only one spectral feature was identified on
    the spectra ($Q=2$).  Shaded regions indicate the HDF and flanking
    fields as in Fig.~\ref{fig-radec-all}.}
\end{figure*}

We show the spatial distribution of objects for which redshifts were
not successfully measured in Fig.~\ref{fig-zfailure}.  Objects which
produced no measurable spectrum ($Q=-2$) because of instrumental
effects are shown as circles.  Since these failures are primarily due
to spectra falling on the gaps between CCDs in the DEIMOS detector
mosaic or to objects placed near the mask edges, they cluster in
several lines across the GOODS-N field.  Targets whose redshifts are
either completely unknown ($Q=1$) or highly uncertain ($Q=2$) are
distributed randomly across the field, as expected.

\begin{figure}[btp]
  \centering
  \includegraphics[angle=0,width=0.475\textwidth]{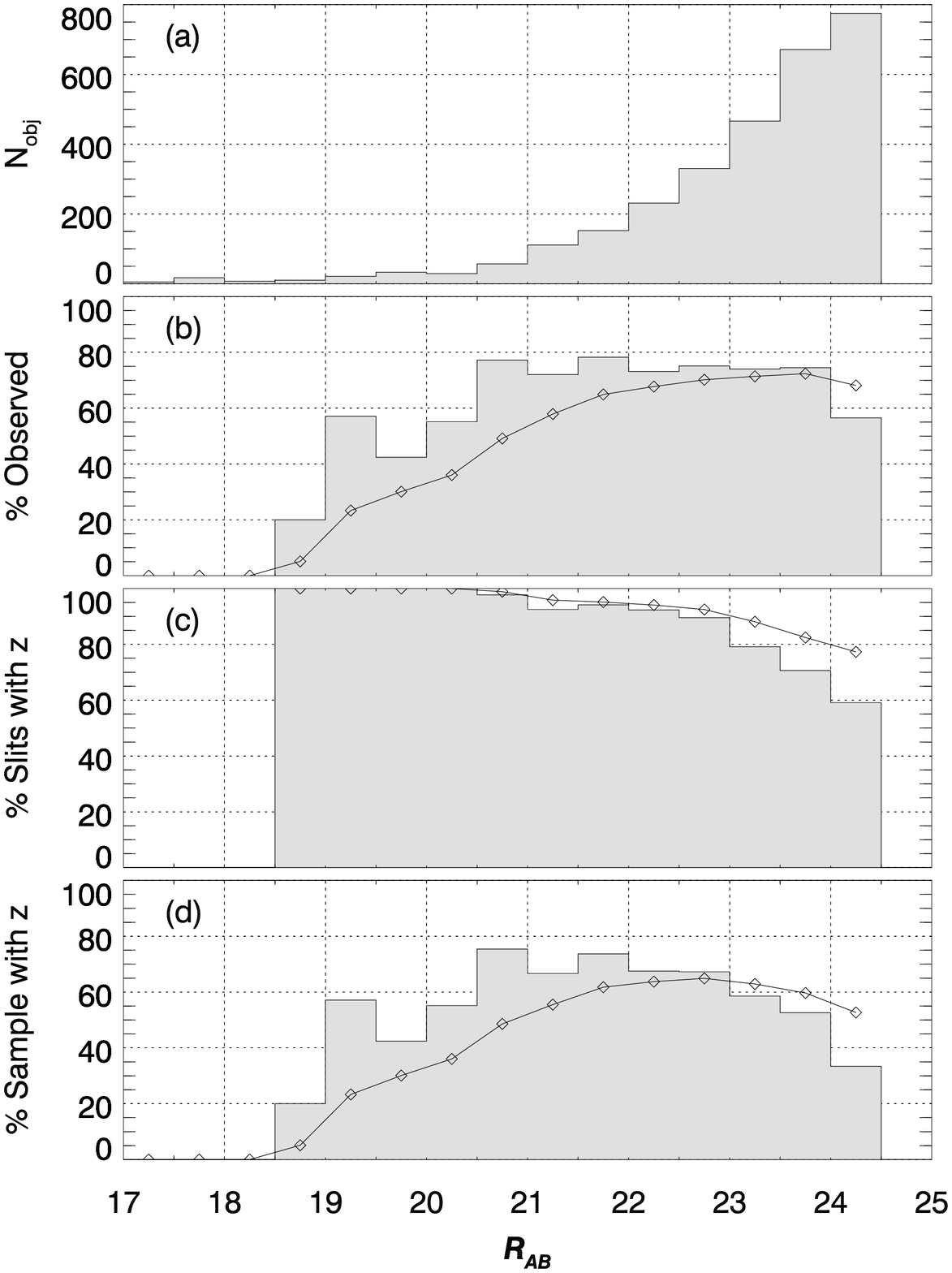}
  \caption{\label{fig-completeness2}
    Completeness indicators as a function of $R_{AB}$ magnitude in the
    TKRS sample.  (a)~The number of available targets $N_{obj}$ per
    0.5~mag bin in our GOODS-N field target catalog, to the limiting
    magnitude of $R\leq24.4$.  (b)~Differential fraction of available
    targets (relative to the total sample available within each
    0.5~mag bin) for which spectra were acquired (\emph{shaded
      histogram}), or the integrated fraction to that magnitude bin
    (\emph{connected diamonds}).  (c)~Differential and integrated
    fraction of \emph{spectra} which yielded secure redshifts.
    (d)~Differential and integrated fraction of \emph{available
      targets} which yielded secure redshifts, including objects not
    observed.}
\end{figure}

\paragraph{Magnitude effects.} Our survey's sampling and success
rates as a function of apparent magnitude are shown in
Fig.~\ref{fig-completeness2} and Table~\ref{tab-completeness}.  For
each 0.5~mag bin, the shaded histogram represents the differential
quantity and the solid line (when present) indicates the integrated
completeness down to that magnitude.
Figure~\ref{fig-completeness2}(a) shows the number of available
targets in our photometric catalog of the GOODS-N field, with the
number in the rightmost bin reduced by the effect of our $R_{AB}=24.4$
cutoff.  Panel (b) shows the fraction of targets we observed
spectroscopically.  Our slitmasks targeted 70--80\% of the available
objects over the range $20.5<R<24$ that contains the majority of the
galaxies, falling significantly below this level at the bright end
(where sources were scarce) and in the faintest bin (containing faint
targets against which selection was weighted during the slitmask
design process).  Panel (c) displays the fraction of these spectra
which yielded secure redshifts (cf. Fig.~4 in \cite{cow04}).  Our
success rate is high ($>90\%$) for galaxies brighter than $R=23.0$ and
quickly decreases to below 60\% at the faint limit.  Panel (d) gives
the fraction of all targets in the sample for which we obtained secure
redshifts, and can be compared to the corresponding figures in
\cite{coh00} and \cite{cow04}.  Overall, our survey obtained secure
redshifts for 53\% of all targets brighter than $R=24.4$ in the
GOODS-N field.

\begin{deluxetable*}{crrrrrrr}[tbp]
  \tablewidth{0pc}
  \tablecaption{TKRS Completeness\label{tab-completeness}}
  \tablehead{
    \colhead{$R_{AB}$ Range\tablenotemark{a}} &
    \colhead{$N_{tot}$\tablenotemark{b}} &
    \colhead{$N_{obs}$\tablenotemark{c}} &
    \colhead{$N_{z}$\tablenotemark{d}} &
    \colhead{$F_{obs}$\tablenotemark{e}} &
    \colhead{$F_{z}$\tablenotemark{f}} &
    \colhead{$F_{z,all}$\tablenotemark{f}} &
    \colhead{$F_{z,cum}$\tablenotemark{g}}\\
    &&&&\colhead{(\%)}&\colhead{(\%)}&\colhead{(\%)}&\colhead{(\%)}}
  \startdata
17.0--17.5 &    5 &    0 &    0 &   0 &   0 &   0 &   0 \\
17.5--18.0 &   17 &    0 &    0 &   0 &   0 &   0 &   0 \\
18.0--18.5 &    7 &    0 &    0 &   0 &   0 &   0 &   0 \\
18.5--19.0 &   10 &    2 &    2 &  20 & 100 &  20 &   5 \\
19.0--19.5 &   21 &   12 &   12 &  57 & 100 &  57 &  23 \\
19.5--20.0 &   33 &   14 &   14 &  42 & 100 &  42 &  30 \\
20.0--20.5 &   29 &   16 &   16 &  55 & 100 &  55 &  36 \\
20.5--21.0 &   57 &   44 &   43 &  77 &  98 &  75 &  49 \\
21.0--21.5 &  111 &   80 &   74 &  72 &  93 &  67 &  56 \\
21.5--22.0 &  152 &  119 &  112 &  78 &  94 &  74 &  62 \\
22.0--22.5 &  231 &  169 &  156 &  73 &  92 &  68 &  64 \\
22.5--23.0 &  330 &  248 &  222 &  75 &  90 &  67 &  65 \\
23.0--23.5 &  466 &  345 &  273 &  74 &  79 &  59 &  63 \\
23.5--24.0 &  671 &  500 &  353 &  75 &  71 &  53 &  60 \\
24.0--24.4 &  775 &  438 &  259 &  57 &  59 &  33 &  53 \\
  \enddata
  \tablenotetext{a}{Apparent magnitude range.}
  \tablenotetext{b}{Number of cataloged objects in magnitude range.}
  \tablenotetext{c}{Number of spectroscopic targets in magnitude range.}
  \tablenotetext{d}{Number of secure galaxy redshifts measured in
    magnitude range.} 
  \tablenotetext{e}{Differential fraction of objects targeted with slits.}
  \tablenotetext{f}{Differential fraction of observed objects yielding secure
    redshifts.} 
  \tablenotetext{f}{Differential fraction of all objects yielding secure
    redshifts.} 
  \tablenotetext{g}{Cumulative fraction of all objects yielding secure
    redshifts.}
\end{deluxetable*}

\begin{figure}[tbp]
  \includegraphics[angle=270,width=0.475\textwidth]{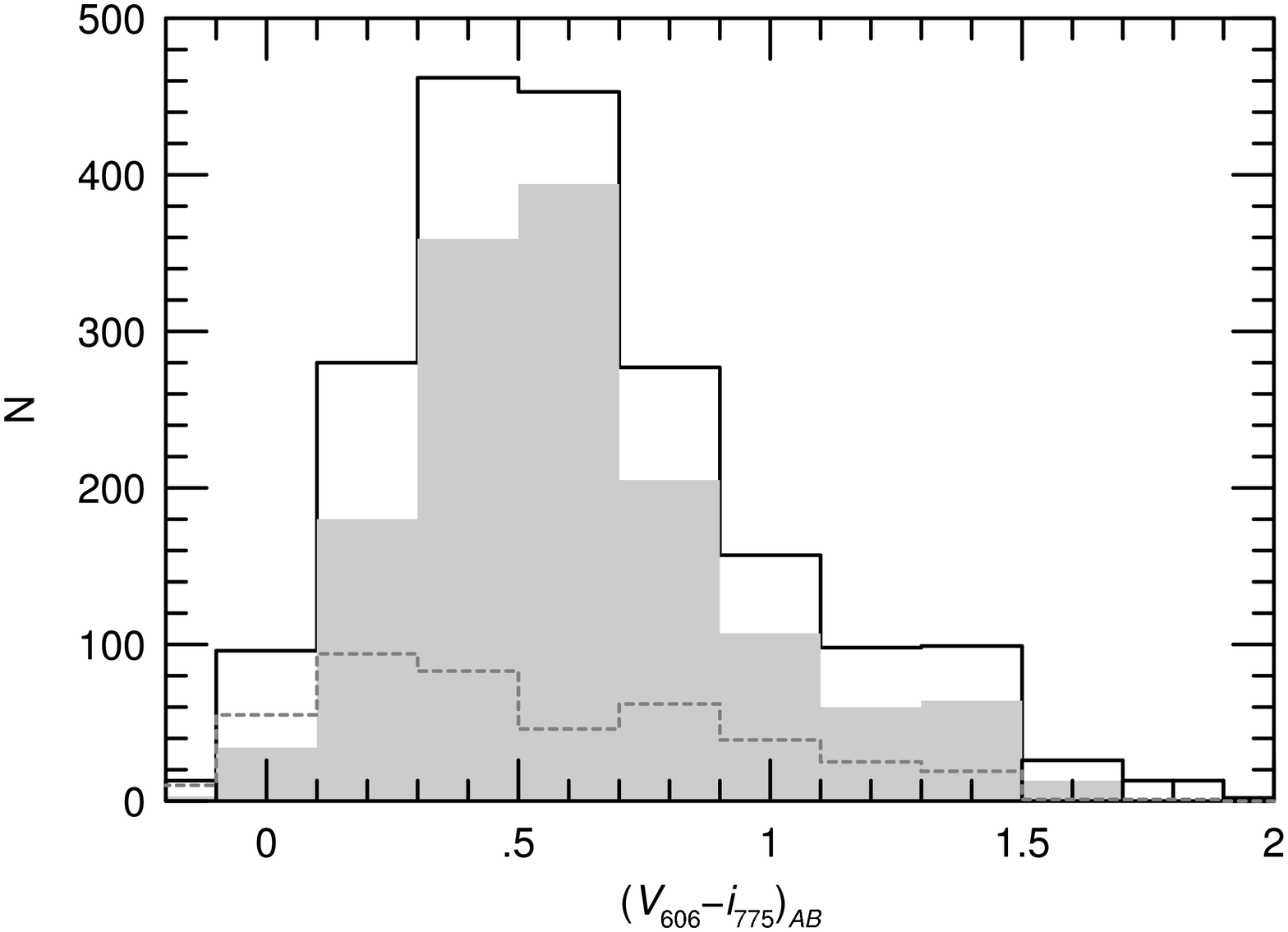}
  \caption{\label{fig-zqual-vi}
    Number of galaxies placed on masks (\emph{solid black line}), and
    number of redshifts measured (\emph{grey histogram}), as a
    function of the \vi\ color.  Also shown is the distribution of
    galaxies with redshift qualities $Q=1$ and 2 (\emph{dashed line}).
    For the bluest galaxies the success rate (ratio of grey histogram
    to solid line) is under 50\%, and, as discussed in the text and
    Fig.~\ref{fig-apparent-cm}, is likely due to unidentified
    high-redshift galaxies.
    }
\end{figure}

\begin{figure}[tbp]
  \includegraphics[angle=0,width=0.475\textwidth]{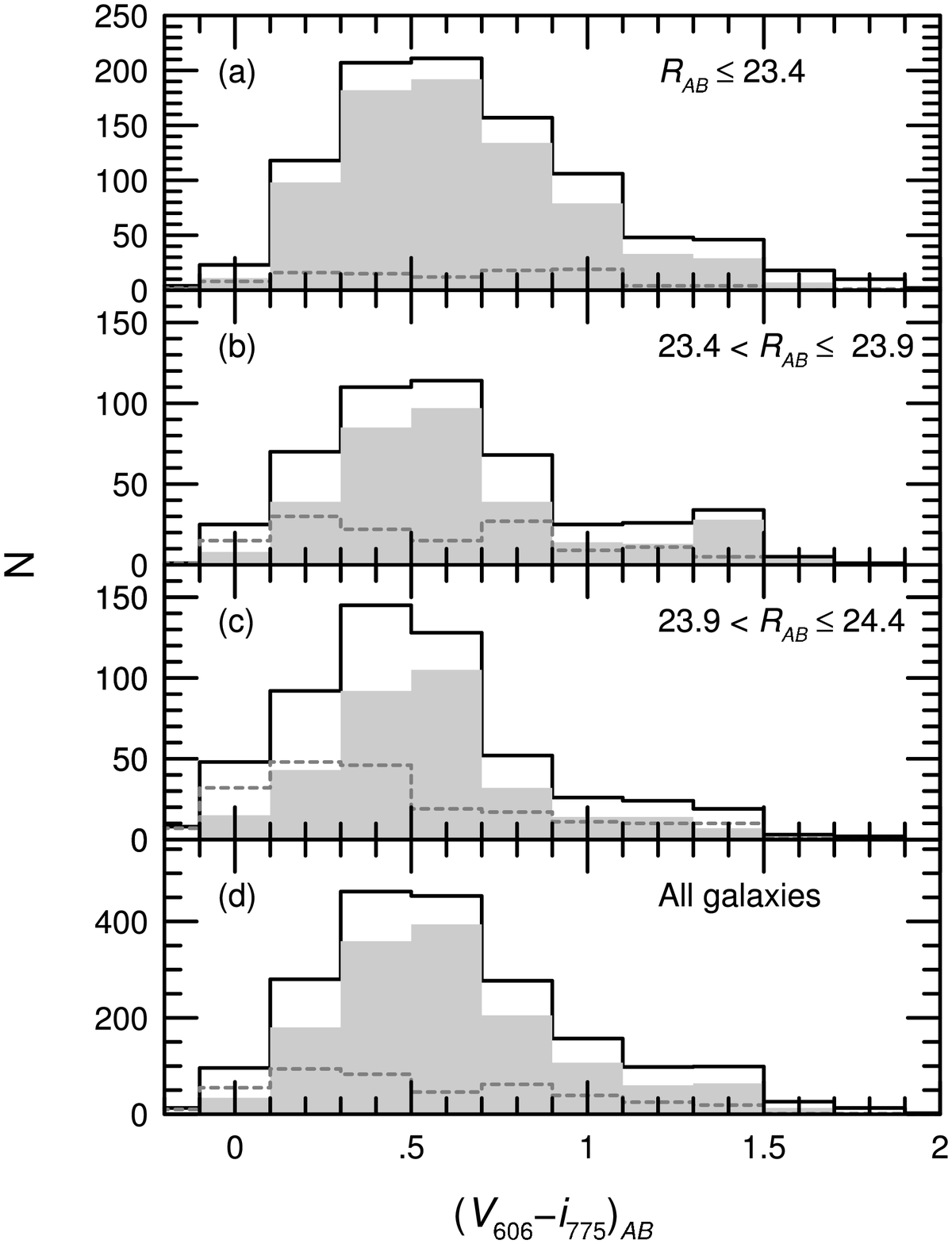}
  \caption{\label{fig-zqual-vi-panels}
    Histogram showing the number of galaxies placed on masks
    (\emph{solid black line}) and number of redshifts measured
    (\emph{grey histogram}), as a function of the $V_{606}~-~i_{775}$
    color.  Also shown is the distribution of failed redshift
    measurements (\emph{dashed line}). Panels (a) through (c) show the
    color distributions for different magnitude ranges, and panel (d)
    shows the distribution for the entire sample. While the color
    distributions of galaxies placed on masks and of those
    successfully observed do not change significantly with magnitude,
    the color distribution of redshift failures shows a clear trend of
    becoming bluer at fainter apparent magnitude.}
\end{figure}

\paragraph{Color effects.} The recent release of multicolor photometry
for the GOODS-N field \citep{cap04,gia04} enables tests of color
dependence in our sample.  In the following analyses, we have matched
our target catalog derived from the DEIMOS photometry with the
Giavalisco \etal\ multicolor photometry derived from the v1.0 GOODS-N
ACS images.  Objects in both catalogs number 2816, with a detection
rate of 2816/2911=90\%. Most non-matching galaxies in the DEIMOS
catalog are found in regions just outside that covered by the v1.0
GOODS-N catalog.  Figure~\ref{fig-zqual-vi} displays the distribution
of the \vi\ colors measured within a circular aperture of radius
$3\arcsec$ for all galaxies placed on mask, galaxies with secure
redshifts ($Q\geq3$) and galaxies with unmeasurable or insecure
redshifts ($Q$=1,2).  The objects with failed redshift
measurements are preferentially bluer than the successful targets.
Figure~\ref{fig-zqual-vi-panels} compares the colors of targets with
successful and failed redshift measurements in different apparent
magnitude ranges, and shows that the distribution of failed
measurements changes from relatively flat to become bluer than the
overall distribution near the faint limit.

\begin{figure*}[tbp]
  \includegraphics[angle=270,width=\textwidth]{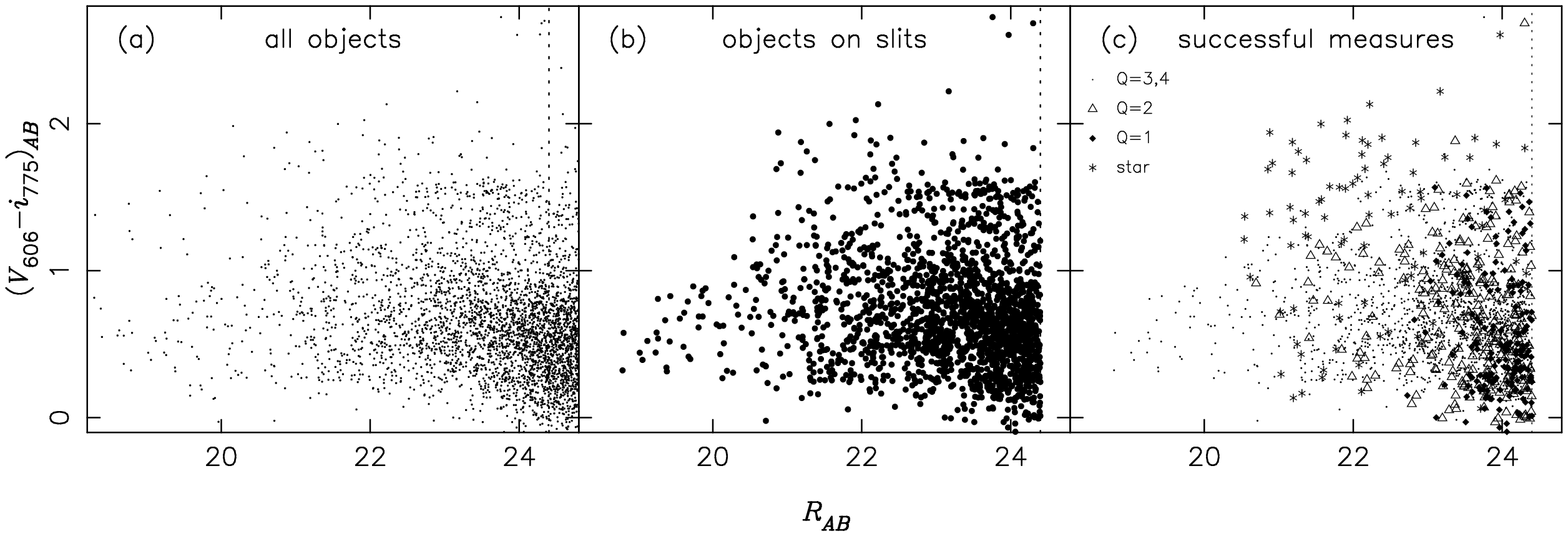}
  \caption{\label{fig-apparent-cm}
    Apparent color-magnitude diagram for objects placed on TKRS masks.
    (a)~The total sample of stars and galaxies in catalog; the dotted
    line represents the nominal limit of $R_{AB}=24.4$ used in the
    selection of the spectroscopic sample. (b)~The distribution of
    objects placed on masks.  (c)~Observed targets, including galaxies
    with secure redshifts ($Q=3,4$), dubious measures ($Q=2$), objects
    for which no redshift can be measured ($Q=1$), and confirmed stars
    ($Q=-1$).  The $Q=1$ sample generally consists of blue and faint
    objects, suggesting that these are likely to be high-redshift
    galaxies for which the [\ion{O}{2}] $\lambda\lambda3726,3729$
    lines fall outside the TKRS spectral window.  }
\end{figure*}

The apparent color-magnitude distribution of the sample in common with
the GOODS-N V1.0 catalogue is shown in Fig.~\ref{fig-apparent-cm},
where panel (a) shows the entire sample (stars and galaxies).  Panel
(b) shows the subsample of objects that were placed on slits, while
panel (c) discriminates targets in various redshift quality classes.
Failed reshifts ($Q=1$) are predominantly found at the faintest and
bluest limits of the sample. A similar analysis made for the DEEP2
survey \citep{wil04}, shows that most of these failed redshifts are
likely to be of galaxies at high redshift, for which the [\ion{O}{2}]
lines fall outside the wavelength range of the DEIMOS setup used in
the TKRS.

\begin{figure*}[tbp]
  \includegraphics[angle=270,width=\textwidth]{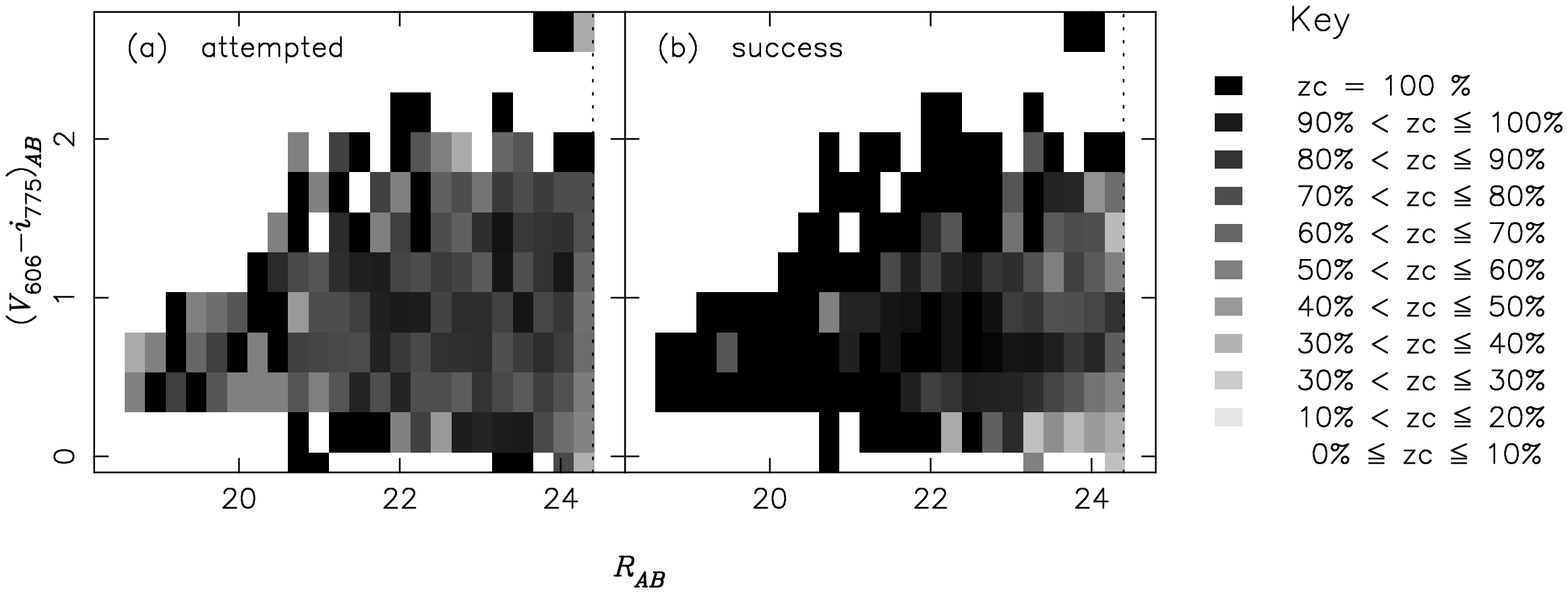}
  \caption{
    Sampling rate for the TKRS sample. (a)~Sampling rate (ratio of
    number of galaxies + stars observed to the number available within
    the bin) vs.\ \HST\ \vi\ color and magnitude, within 0.25~mag bins.
    (b)~Success rate (ratio between the number of successful
    measurements and the number placed on masks) vs.\  \HST\ \vi\ color
    and magnitude.  The legend to the right of the figure explains the
    color coding.
    \label{fig-sampling}
    }
\end{figure*}

The sampling rate as a function of apparent \vi\ color and magnitude
is shown in Fig.~\ref{fig-sampling}. The sampling rate was measured in
bins of 0.25~mag both in magnitude and color, and the shading in the
figure represents different sampling rates as described in the key.
Panel (a) shows the ratio between number of objects placed on masks,
relative to the number of objects within the same color-magnitude
limits in the sample. The sampling rate of the TKRS is higher than 65\%
for the majority of the color-magnitude bins, though this rate falls
for fainter and bluer galaxies. Panel (b) shows the success rate; \ie,
the number of spectra that produced redshifts compared to the total
placed on masks. The completeness level is higher than 80\% to about
$R\la23$, beyond which it falls due to the lower $S/N$ of absorption
line spectra, and the loss of key spectral features outside the
wavelength range of the spectra.

\begin{figure}[tbp]
  \includegraphics[angle=270,width=0.475\textwidth]{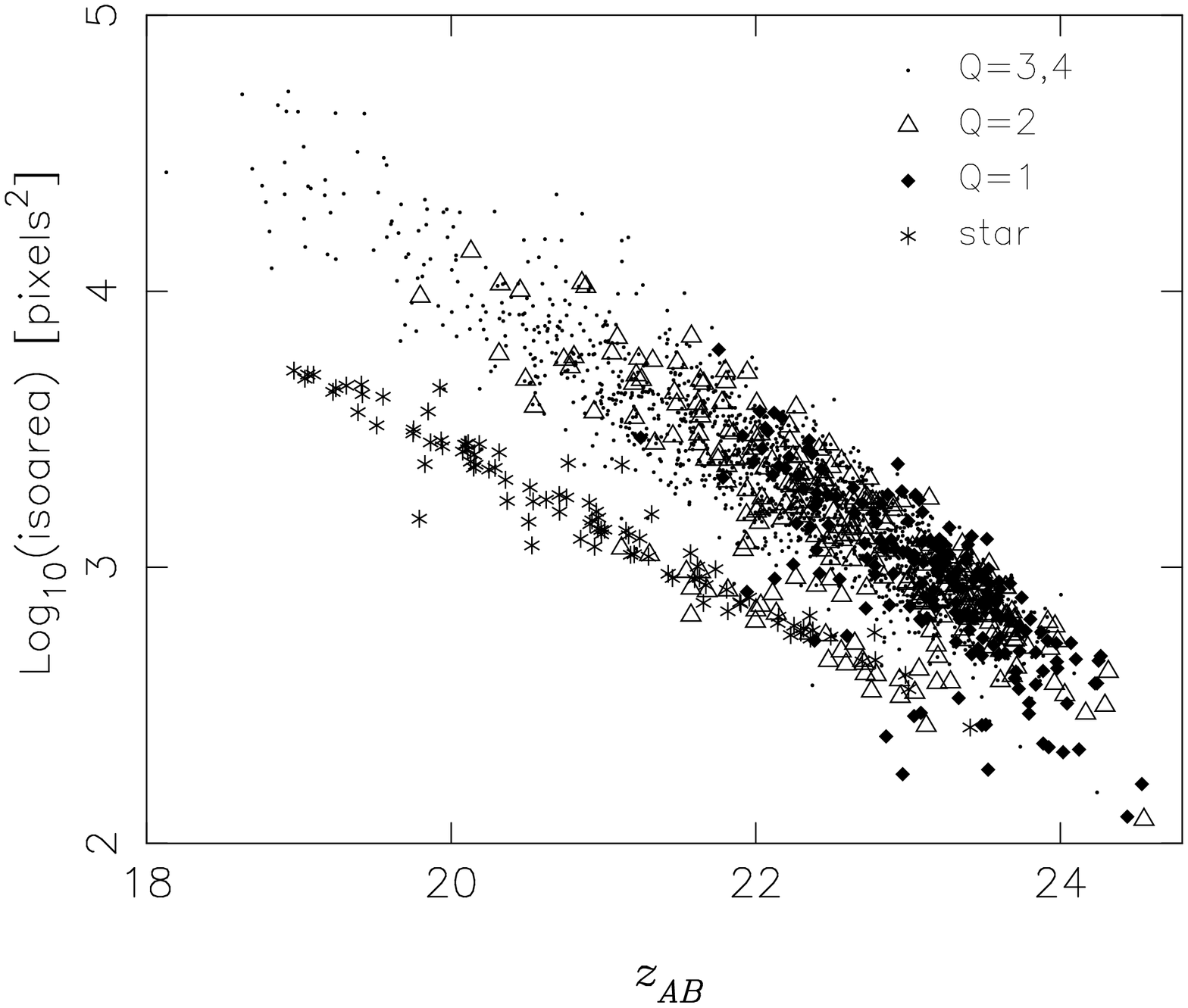}
  \caption{\label{fig-zmag-area}
    Apparent object size (ACS isophotal image area) vs.\ apparent $z$
    magnitude measured in the outermost aperture ($3\arcsec$) on ACS
    images \protect\citep{gia04} for the 1981 sources observed in our redshift
    survey.  Different symbols distinguish objects with secure redshifts
    ($Q=3,4$), uncertain redshifts ($Q=2$), no measured redshifts
    ($Q=1$), and stars ($Q=-1$).}
\end{figure}

\paragraph{Surface brightness effects.}
Figure~\ref{fig-zmag-area} shows isophotal area plotted against $z$
magnitude for all of the spectroscopic targets in our survey.  Both
photometric parameters are from the \HST\ ACS imaging catalog of
\cite{gia04}, with the magnitudes measured in the outermost aperture
($3\arcsec$).  In this plot we again distinguish objects in the
various redshift quality classes of our survey.  The stellar and
galaxy sequences are well differentiated to the limiting $z$
magnitude.  Objects with uncertain redshifts include faint stars as
well as galaxies of varying brightness.  Many of the brighter objects
with uncertain redshifts ($Q=2$) failed because of poorly defined
extraction windows when creating the 1-D spectra.  This problem will
be revisited in a later re-analysis of the 2-D spectra and a corrected
version of the catalog will be issued.  Failed redshifts ($Q=1$), as
opposed to uncertain ones, start to occur at $z\approx22$.  However,
the diagram provides no strong evidence that the $Q=1$ galaxies have
lower surface brightnesses than the subsample with secure redshifts.

\subsection{Comparison with other surveys}\label{sxn-other-surveys}

Areas within the GOODS-N field have already been the subject of
intensive study by deep surveys of the HDF-N region that lies within
it.  As mentioned in \S\ref{sxn-intro}, several research teams have
obtained redshifts for galaxies in this part of the sky, among them
\cite{coh96,cow96,ste96,low97,phi97,mou97,coh00,coh01,daw01}.  Recent
work by other observing teams using Keck~II with DEIMOS to survey the
GOODS-N field include the companion survey of \cite{cow04} and the
higher-resolution survey of red galaxies by \cite{tre04}.  We can
compare our measurements with the findings of these independent
surveys to gauge the accuracy of our redshifts and our quality
assessments.

\begin{figure}[!btp]
  \centering
  \includegraphics[angle=0,width=0.475\textwidth]{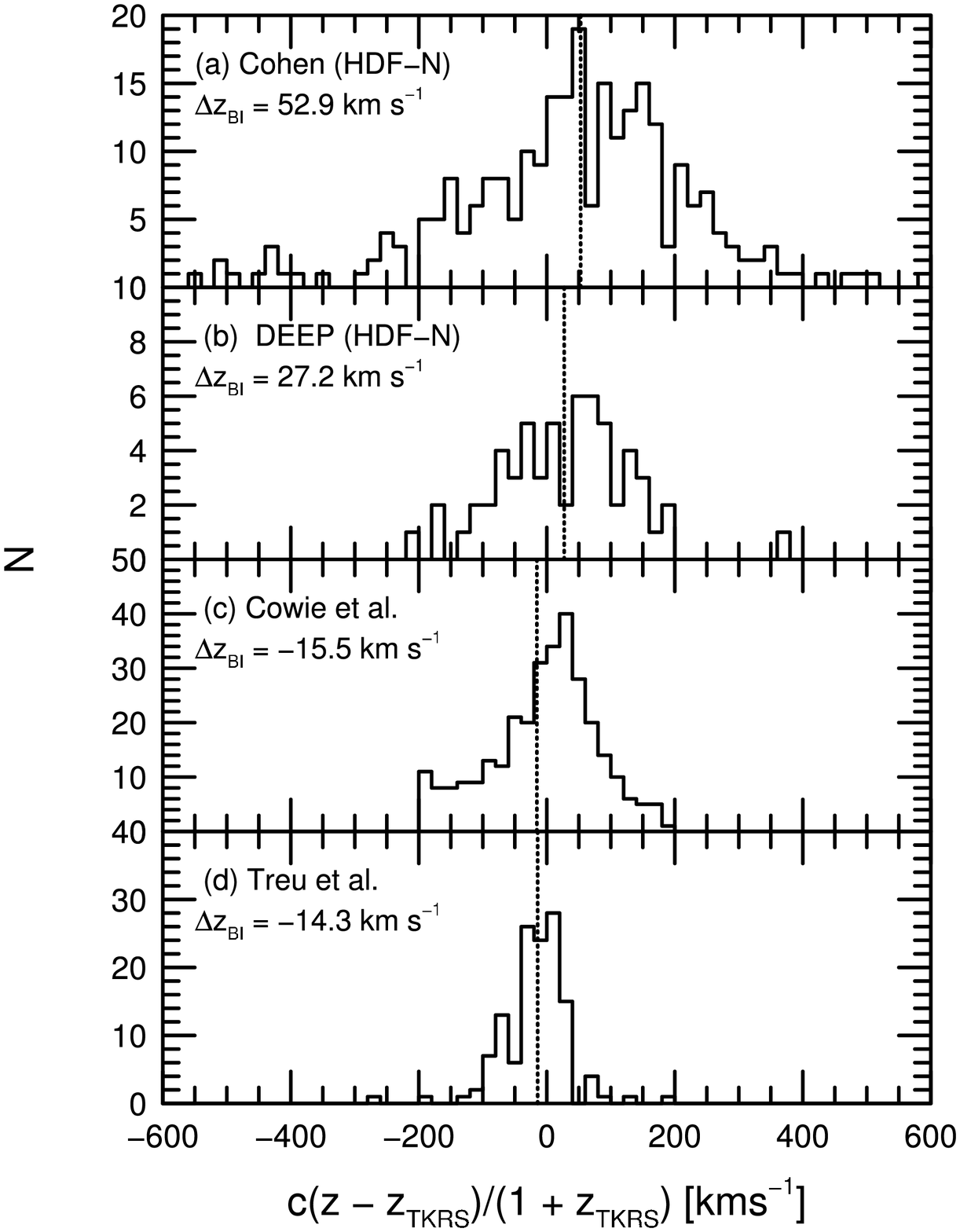}
  \caption{\label{fig-zhist}
    Histogram showing the redshift differences between galaxies with
    secure measurements in the TKRS vs.\ four other extensive samples
    in the GOODS-N region. Each panel shows the bi-weight location
    (\emph{dotted line}) and its value \protect\citep{bee90}.  (a)~Redshift
    differences for galaxies with secure measurements from LRIS in
    \protect\cite{coh00} and \protect\cite{coh01}.  (b)~Redshift differences between
    TKRS and secure measurements of \protect\cite{phi97} and \protect\cite{low97},
    also using LRIS.  (c)~Redshift differences with the \protect\cite{cow04}
    sample of galaxies, which used similar instrumental setup and
    integration times as the TKRS.  (d)~Redshift differences with the
    \protect\cite{tre04} sample of galaxies observed with DEIMOS but using
    higher resolution and longer integration times than this work.}
\end{figure}

The results of the redshift comparisons are depicted graphically in
Fig.~\ref{fig-zhist} and key results are summarized in
Table~\ref{tab-zcompare}.  For each survey, the table provides the
number of objects in common with the TKRS sample in addition to robust
measures of both the difference in measured redshift and the
dispersion in that difference.  For purposes of this discussion we
defined two redshift measurements to ``agree'' if they differ by less
than $|\Delta z|\le0.002$ ($|\Delta v|\la600$~\kms); this excludes
galaxies with severely discrepant redshifts and does not significantly
bias the resulting measures of redshift offset and dispersion.  To
quantify the redshift differences with our work we computed the
offsets via the biweight, while the median absolute deviation from the
median (MAD, with multiplicative scaling factor 9) served as a robust
estimator of the dispersion in the offsets \citep{bee90}.  We now
discuss the results of the comparison for individual surveys.

\begin{deluxetable}{lcrrr}[btp]
  \tablewidth{0pc}
  \tablecaption{Comparison of TKRS with Other
    Surveys\label{tab-zcompare}}
  \tablehead{
    \colhead{Survey}                  &
    \colhead{Ref.}               &
    \colhead{$N_{obj}$\tablenotemark{a}}               &
    \colhead{$\Delta v_{BI}$\tablenotemark{b}}  &
    \colhead{$S_{MAD}(\Delta v)$\tablenotemark{c}} \\
    &&&(\kms)&(\kms)}
  \startdata
  Cohen \etal\ HDF-N   & 1,2 & 254 &    52.9 & 174.3\\ 
  DEEP1 HDF-N          & 3,4 &  60 &    27.2 &  99.8\\ 
  Cowie \etal\ GOODS-N & 5 & 378 & $-15.5$ & 140.5\\ 
  Treu \etal\ GOODS-N  & 6 & 131 & $-14.3$ &  42.2\\  
  \enddata
  \tablerefs{
    (1)~\protect\cite{coh00};
    (2)~\protect\cite{coh01};
    (3)~\protect\cite{low97};
    (4)~\protect\cite{phi97};
    (5)~\protect\cite{cow04};
    (6)~\protect\cite{tre04}.}
  \tablenotetext{a}{Number of objects with secure redshifts in both surveys.}
  \tablenotetext{b}{Difference in radial velocity (in the sense
    $v-v_{TKRS}$) among secure redshifts, as quantified by the Biweight.}
  \tablenotetext{c}{Robust estimate of dispersion in radial velocity
    differences as quantified by the median absolute deviation from
    the median, scaled by an 
    appropriate factor \protect\citep{bee90}.}
\end{deluxetable}

In an LRIS survey covering a region centered on the HDF and Flanking
Fields, Cohen \etal\ obtained spectra for most galaxies with
$R\leq23.5$ in the Flanking Fields and $R\leq24.5$ in the HDF-N
\citep{coh96,coh00,coh01}.  The redshift catalogue of these works also
contains measurements obtained from the literature; however, in the
present comparison we concentrate on the uniform sample of 413
galaxies measured by J.~Cohen using Keck~I with LRIS.  For 254
galaxies with redshift in rough agreement with ours
(Fig.~\ref{fig-zhist}(a)) the biweight difference is $z_{Cohen} -
z_{TKRS} = 53$~\kms\ with a dispersion of 174~\kms.  \cite{coh00}
obtained secure redshifts for 52 targets measured poorly in the TKRS
sample ($Q=1$ or 2).

Another LRIS sample in the HDF-N and Flanking Fields is that observed
by the DEEP1 survey team, who targeted primarily $U$- and $B$-band
dropouts \citep{low97} and compact galaxies \citep{phi97}. Of 114
galaxies in common, 60 have redshifts within $|\Delta z|<0.002$ of the
TKRS measurement, and the distribution of redshift differences is shown
in Fig.~\ref{fig-zhist}(b). The offset in redshift is $z_{DEEP} -
z_{TKRS} = 27$~\kms\ with a dispersion of 100~\kms.  The DEEP1 sample
provides 7 reliable measurements of galaxy redshifts for which the TKRS
redshift has a quality of 1 or 2.

The \cite{cow04} sample, which used the same instrument,
configuration, integration time, and survey region as the TKRS sample,
features 830 galaxies in common with TKRS, of which 452 have secure
redshifts in both works; see Fig.~\ref{fig-zhist}(c).  Among the 378
redshifts in agreement, the offset is $z_{Cowie} - z_{TKRS} =
-16$~\kms\ with a dispersion of 141~\kms.  A total of 56 galaxies for
which TKRS was unable to obtain a redshift (\ie, $Q=1,2$) were
successfully measured by Cowie \etal\ Conversely, TKRS was able to
measure redshifts for 50 galaxies that were observed by Cowie \etal\ 
but for which they were unable to get a reliable redshift.

Finally, in Fig.~\ref{fig-zhist}(d) we compare our redshifts with
measurements obtained by \cite{tre04}, who targeted preferentially red
galaxies in the GOODS-N field using Keck~II with DEIMOS.  Their
higher-resolution survey used the 1200~l~mm\inv\ grating to acquire
exposures ranging from 5--10~h, as compared to 1~h for TKRS.  Of 199
galaxies in common with our catalog, 178 have secure redshifts in the
Treu \etal\ sample and 131 have secure redshifts in both samples; only
one redshift measurement is significantly discrepant ($| \Delta z
|\geq0.0025$).  Among 130 redshifts which agree the offset between
these samples is $-14$~\kms\ with a dispersion of 42~\kms, so that
this sample represents the lowest dispersion with respect to our
redshifts.  These measurements from Treu \etal\ also provide an
estimate on how the increase of exposure time affects the redshift
success rate. Of the 18 objects in common whose redshifts in the TKRS
were uncertain ($Q=2$) or could not be measured, 7 (39\%) agree with
the measurements with longer exposures.  This number corresponds to
$\sim4\%$ (7/179) of galaxies within the sample in common. An
additional 6\% (11/179) of objects yielded successful redshift
measurements with the longer integration times.

\subsection{Accuracy of redshift measurements}

The independent redshift measurements available for a significant
number of galaxies in our catalog allow us to estimate the uncertainty
in our redshifts.  Of the four comparable catalogs described in
\S\ref{sxn-other-surveys}, the most useful ones for comparison are the
DEIMOS surveys of Treu \etal\ and Cowie \etal, since they use the same
instrument as our survey.  Of these two, the Treu \etal\ sample is
expected to be more accurate by virtue of its higher dispersion (twice
that of TKRS) and longer exposure time (typically 5 times longer).  The
dispersion in redshift measurements between these two surveys should
thus be dominated by errors in the TKRS sample.  As shown in the last
column of Table~\ref{tab-zcompare}, the robust estimate of dispersion
in the redshift differences of these surveys is 42~\kms.  Assuming
that the TKRS velocity uncertainties are at least a factor of two
larger than those of Treu \etal\ due to the difference in dispersion,
we estimate an average uncertainty of about 38~\kms\ for TKRS, with the
caveat that the sample of galaxies used in the comparison consists
almost exclusively of early-type galaxies and is thus not
representative of the GOODS-N region as a whole.

The most comparable of the existing surveys to TKRS is the Cowie
\etal\ sample made at the same dispersion and wavelength interval with
DEIMOS.  The dispersion in redshift measurements for the sample of 363
common targets between these two surveys is 141~\kms.  Assuming the
two surveys have comparable uncertainties, this implies an error of
99~\kms\ for each.  This value is over a factor of two higher than
that measured compared to the Treu \etal\ sample.  This may be due to
Cowie \etal\ surveying fainter galaxies over a wider range of spectral
types than Treu \etal, as well as different slit orientations used by
Cowie \etal\ vs.\ TKRS.  Being conservative, we conclude that the
characteristic uncertainty in our redshift measurements is
$\sigma\approx100$~\kms\ or better.

\subsection{Accuracy of Quality Codes}\label{sxn-zquality}

The existence of independent redshift measurements for many objects in
our catalog allows us to assess the accuracy of our redshift quality
categories.  In other words, are the quality code 3 redshifts (which
by definition should be correct at the 95\% confidence level) truly
correct 95\% of the time?  In Table~\ref{tab-zqual2} we match the TKRS
catalogue with a list of redshifts measured by other authors. This
list collated measurements from
\cite{coh96,cow96,low97,phi97,mou97,coh00,coh01,daw01,ste03} and
\cite{cow04}.  The collated catalog was then correlated against the
DEIMOS catalog, providing a list of matching objects.  In assessing
the number of correct redshift identifications, we assume that all
discrepant redshifts result from errors in the present sample, thus
obtaining a firm \emph{lower limit} on the quality of our
measurements.  The final column in this table shows that the $Q=4$
redshifts agree with previous measurements in 94\% of cases,
establishing this as the lower limit to the accuracy of our most
secure redshifts. For $Q=3$, the corresponding lower limit on the
success probability is 91.6\%.  If we make the more reasonable
assumption that the mis-identifications are equally split between TKRS
and other samples, the success rate increases to 96\% ($Q=3$) and 97\%
($Q=4$).  We conclude that the respective intended confidence levels
of 95\% and 99\% for the category 3 and 4 redshift measurements are
reasonably accurate.

\subsection{Comparison with photometric redshifts}

\begin{figure}[tbp]
  \includegraphics[angle=90,width=0.475\textwidth]{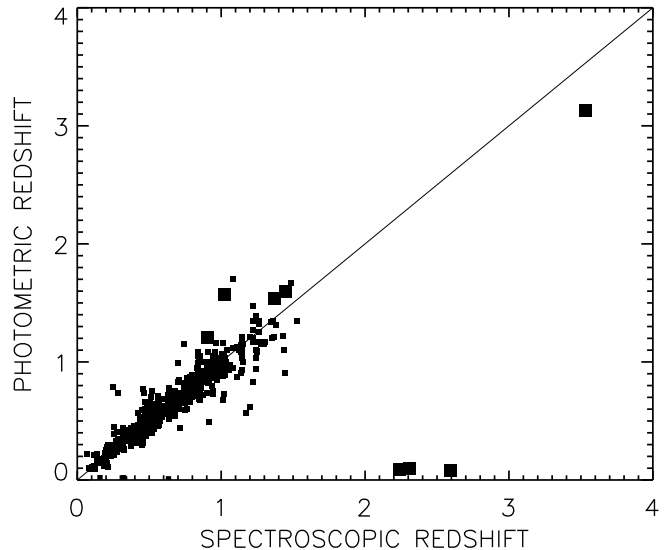}
  \caption{\label{fig-photo-z}
    Photometric redshift versus spectroscopic redshift for the 1078
    sources with $R<24$ where both have been measured (\emph{solid
      squares}).  Objects with broad-line AGN spectra are
    distinguished with a larger symbol. Only 38 of the photometric
    redshifts differ by more than 30\% from the spectroscopic
    redshift.  Five of these cases are broad-line AGN.}
\end{figure}

We have compared the spectroscopic redshifts with the photometric
redshift catalog of \cite{cap04}.  These redshift estimates are based
on the ultradeep imaging in eight colors from the U to the near IR
which is described in \cite{cap03} and are computed using the Bayesian
code of \cite{ben00}.  This assigns odds that the redshift lies close
to the estimated value and we have only used those cases where these
odds lie above 90\%. The imaging data also saturate at bright
magnitudes so we have restricted the comparison to sources with
$R>20$. This gives a sample of 1078 sources with both spectroscopic
and photometric redshifts which we compare in Fig.~\ref{fig-photo-z}.
Of the sources, 8 are broad line AGN where the photometric estimates
may be poorer since the templates do not properly model such objects.
We have distinguished these objects with larger symbols.  Nearly all
(97\%) of the photometric redshifts agree to within 30\% with the
spectroscopic redshift, consistent with the expectation from the
probabilities assigned to the photometric redshifts. This agreement is
independent of magnitude over the $20<R<24$ range.

\section{Analysis}\label{sxn-analysis}

\subsection{Redshift distributions}

\begin{figure}[!btp]
  \includegraphics[angle=90,width=0.475\textwidth]{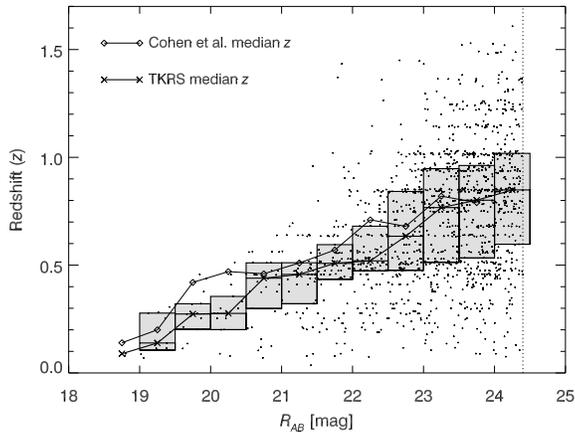}
  \caption{\label{fig-zmedian}
    Redshift vs.\ $R$ magnitude for targets with secure redshifts.
    Small pluses indicate individual objects in the TKRS survey.
    Boxes indicate the interquartile range of redshifts within
    respective 0.5~mag bins; each has a horizontal line and $\times$ which
    indicate the median redshift within that bin.  The median
    redshifts of \protect\cite{coh00} are shown for comparison.  The vertical
    dotted line at right represents the apparent magnitude limit of
    the present survey at $R_{AB}=24.4$.}
\end{figure}

The distribution of galaxies in our survey as a function of $R_{AB}$
and redshift appears in Fig.~\ref{fig-zmedian}.  Also shown in this
plot and in Table~\ref{tab-zmedian} is the median redshift as a
function of magnitude.  The plot compares these values to the median
redshifts in the \cite{coh00} survey; both samples are uncorrected for
incompleteness.  For bins brighter than $R=22.5$, our median redshifts
are significantly lower than those of Cohen \etal.  We note that the
present survey has a cumulative completeness of just over 60\% at
$R=22.5$, compared to $>90\%$ for Cohen \etal.  In the faintest
magnitude bin of the Cohen \etal\ HDF sample (23.5--24.0) the
differential completeness drops to the same 50\% value as our survey,
and as a result the median redshifts in this magnitude range
($z_M=0.79$ for Cohen \etal\ vs.\ 0.80 in our sample) are consistent.

\begin{deluxetable}{crccc}[tbp]
  \tablewidth{0pc}
  \tablecaption{Median Redshift in Bins of $R$\label{tab-zmedian}}
  \tablehead{
    \colhead{$R_{AB}$ Range\tablenotemark{a}} &
    \colhead{$N_{obj}$\tablenotemark{b}} &
    \colhead{Q1\tablenotemark{c}} &
    \colhead{Median\tablenotemark{d}} &
    \colhead{Q3\tablenotemark{e}}}
  \startdata
  18.5--19.0 &    2 &  \nodata & 0.09 &  \nodata \\
  19.0--19.5 &   11 & 0.11 & 0.14 & 0.28 \\
  19.5--20.0 &   13 & 0.20 & 0.27 & 0.32 \\
  20.0--20.5 &   15 & 0.20 & 0.28 & 0.36 \\
  20.5--21.0 &   33 & 0.30 & 0.44 & 0.51 \\
  21.0--21.5 &   57 & 0.32 & 0.46 & 0.51 \\
  21.5--22.0 &   93 & 0.43 & 0.51 & 0.60 \\
  22.0--22.5 &  133 & 0.47 & 0.52 & 0.68 \\
  22.5--23.0 &  193 & 0.48 & 0.63 & 0.84 \\
  23.0--23.5 &  254 & 0.51 & 0.77 & 0.95 \\
  23.5--24.0 &  335 & 0.53 & 0.80 & 0.96 \\
  24.0--24.4 &  248 & 0.60 & 0.85 & 1.02 \\
  \enddata
  \tablenotetext{a}{Apparent magnitude range.}
  \tablenotetext{b}{Number of secure TKRS redshifts in magnitude range.}
  \tablenotetext{c}{Lower quartile redshift within magnitude range.}
  \tablenotetext{d}{Median redshift within magnitude range.}
  \tablenotetext{e}{Upper quartile redshift within magnitude range.}
\end{deluxetable}
    
\begin{figure*}[ptb]
  \includegraphics[angle=90,width=\textwidth]{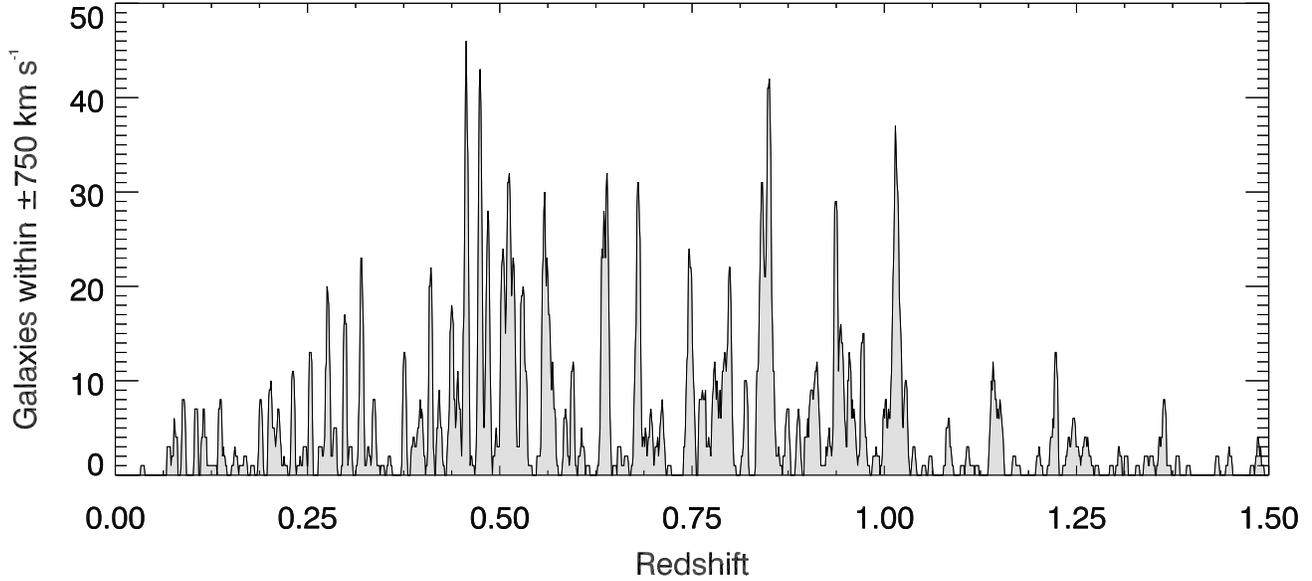}
  \caption{
    \label{fig-gdwzhist}
    Redshift distribution of galaxies with secure measurements in the
    present sample.  The raw marginal distribution was smoothed with a
    moving top hat kernel tuned to highlight structures similar in
    velocity dispersion to those noted in the HDF \protect\citep{coh00}.  The
    ordinate represents the number of galaxies within $\pm750$~\kms\ 
    at any given redshift.  Several galaxies at high redshift are not
    shown.}
\end{figure*}

\begin{figure*}[ptb]
  \includegraphics[angle=90,width=\textwidth]{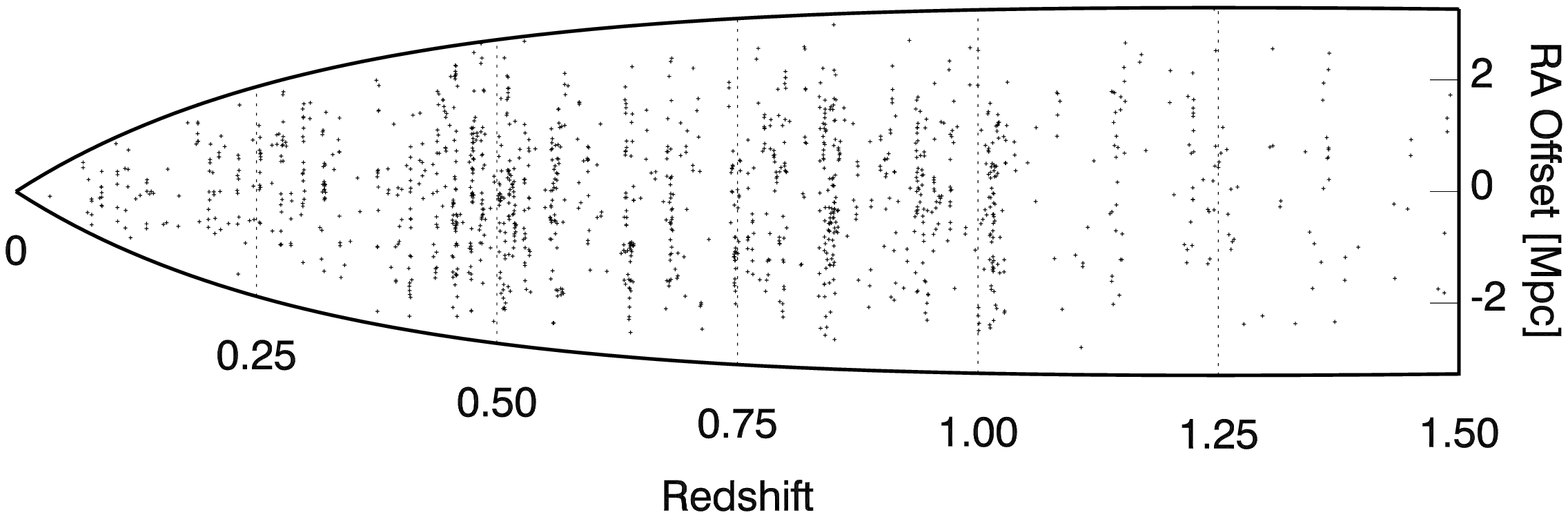}\\
  \vspace*{1em}\\
  \includegraphics[angle=90,width=\textwidth]{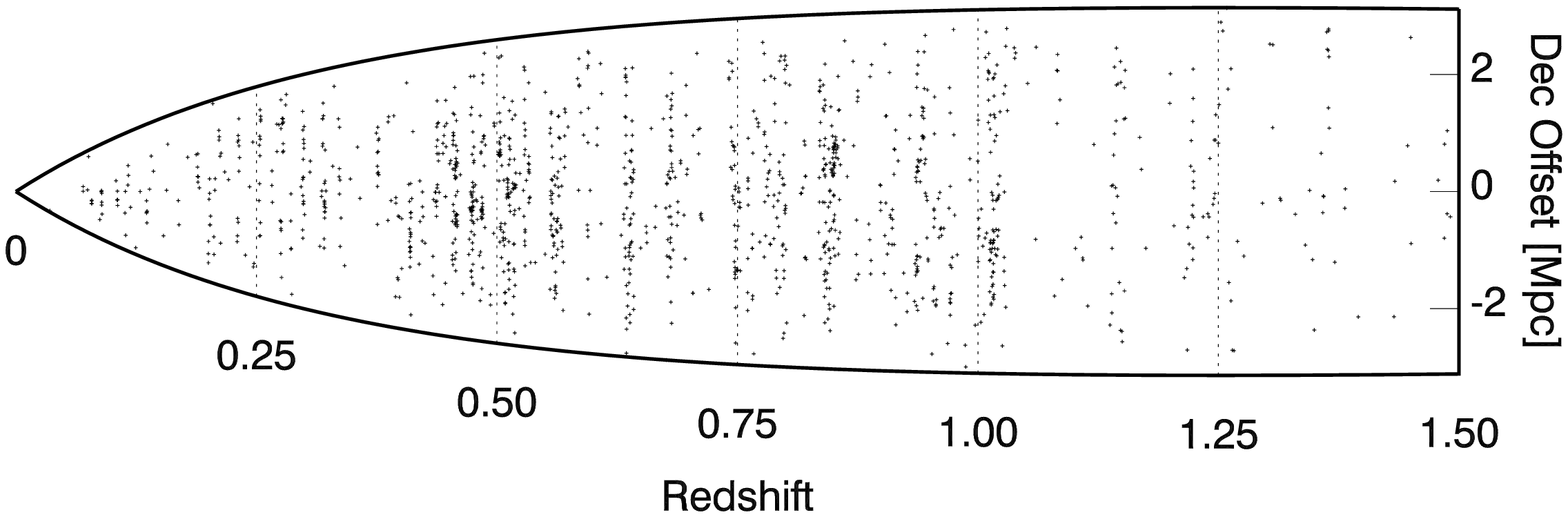}
  \caption{\label{fig-pie}
    Pie diagrams showing the spatial distribution as a function of
    redshift for all galaxies with secure measurements in our survey.
    (a)~Projected distance of each galaxy from the center of the
    GOODS-N field in the direction of Right Ascension vs.\ redshift.
    (b)~Same, for Declination.  In each plot, the outer envelope
    represents the linear separation corresponding to the $18\farcm4$
    width of the field at that redshift.  Note the numerous walls
    correponding to peaks in the marginal distribution of redshifts
    seen in Fig.~\ref{fig-gdwzhist}.  Dotted lines indicate lines of
    constant redshift.  Cosmological parameters $h_0=0.75$ and
    $q_0=0.5$ are assumed in computing the spatial offsets.}
\end{figure*}

Figure~\ref{fig-gdwzhist} shows the redshift distribution of galaxies
with secure measurements.  This plot shows the number of galaxies
contained within a velocity window of width $\pm750$~\kms\ at any
given redshift and thus isolates concentrations of galaxies in
redshift space.  The window width of 1500~\kms\ is well-matched to the
numerous structures with velocity dispersions of
$\sigma=$~500--700~\kms\ reported by \cite{coh00} in the HDF.
Numerous concentrations of galaxies are seen in the sample, most
notably those at $z=0.51$, 0.56, 0.64, 0.85, and 1.02.  These
structures are equally prominent in the pie diagrams shown in
Fig.~\ref{fig-pie}(a) and (b) which plot the projected linear offset
from the field center (along the $\alpha$ and $\delta$ axes,
respectively) vs.\ redshift.  These figures illustrate that some of
the structures span the GOODS-N field, while others (particularly
those at $z=0.85$ and 1.02) are spatially clumped.

\subsection{Color distribution}

\begin{figure}[tbp]
  \centering
  \includegraphics[angle=0,width=0.475\textwidth]{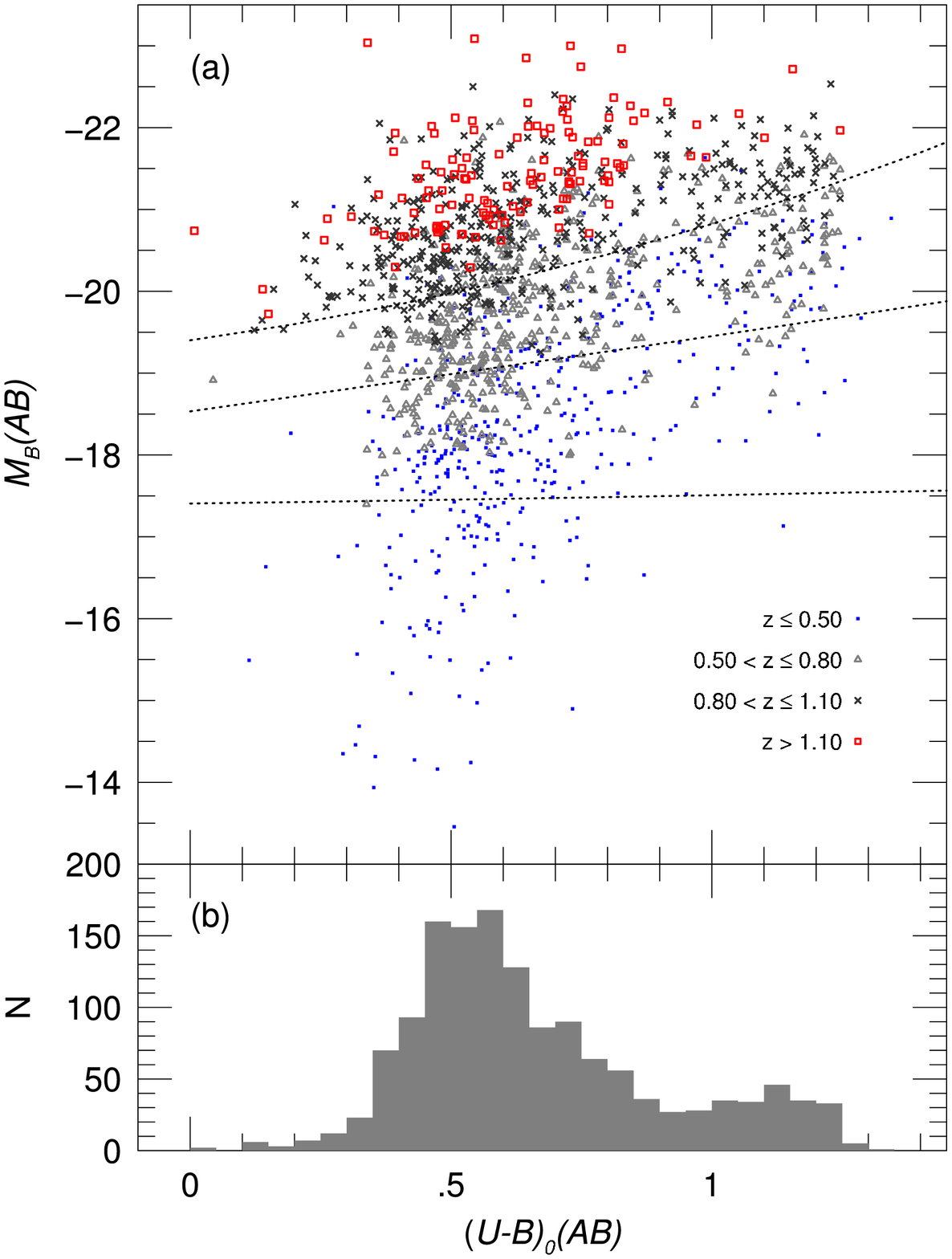}
  \caption{\label{fig-absolute-cmd}
    Rest-frame color-magnitude diagram and color distribution.
    (a)~Rest-frame brightness vs.\ color for galaxies with TKRS
    redshifts in the \protect\cite{gia04} catalog.  The three dashed lines
    show the limiting absolute magnitude as a function of restframe
    color at three different redshifts: $z=0.50$ (\emph{lower curve}),
    $z=0.80$ (\emph{middle curve}) and $z=1.10$ (\emph{top curve}).
    The symbols are also coded by redshift interval as shown in the
    figure.  (b)~Distribution of $(U-B)_0$ colors in 0.05~mag bins.
    The trough at $(U-B)_0(AB) \approx 0.9$ is consistent with that
    seen in other samples of galaxies.  }
\end{figure}

In Fig.~\ref{fig-absolute-cmd} we show the rest-frame color-magnitude
diagram based on the \HST\ ACS photometry for all galaxies with secure
TKRS redshifts appearing in the \cite{gia04} catalog.  The rest-frame
parameters were calculated by convolving the CCD and filter
throughputs of ACS with spectral energy distributions of galaxies in
the \cite{kin96} atlas, a more detailed description of which will be
presented elsewhere \citep{wil04}.  As seen in
Fig.~\ref{fig-absolute-cmd}(b), the rest-frame galaxy colors are
bi-modal about a minimum near $(U-B)_0(AB) \approx 0.9$ (corresponding
to about $(U-B)_0 \approx 0.3$ in the Vega system).  This effect has
also been noted in the DEEP1 survey \citep{im02,wei04} in the Groth
Survey Strip (which also used \HST\ photometry), as well as in the Sloan
Digital Sky Survey \citep{str01,hog02} and COMBO-17 survey \citep{bel03}.
Figure~\ref{fig-absolute-cmd}(a) illustrates that the the color
bi-modality is clearly present to $z > 1.1$, which is beyond the
limits both of \cite{bel03} and \cite{wei04}.  The lack of
low-luminosity red galaxies, also noted by \cite{wei04}, is most
likely to be a real effect.

\section{Summary}\label{sxn-summary}

We have described our deep redshift survey of the GOODS-N region using
DEIMOS, the result of which is a catalog of photometry, astrometry,
redshifts and redshift quality assessments for 1440 confirmed
galaxies/AGN and stars in the $10\arcmin\times16\arcmin$ GOODS-N
region.  This sample is 53\% complete to a limiting magnitude of
$R_{AB}=24.4$ and has a median redshift of $z_M=0.65$.  In conjunction
with an accompanying DEIMOS survey by \cite{cow04}, it provides a
substantial and homogeneous sample of new redshifts and spectra in
this important area of the sky.

The TKRS catalog will continue to evolve as DEIMOS pipeline processing
algorithms are refined, and we expect that a significant number of
galaxies may be promoted from redshift quality class 2 (uncertain) to
3 ($>90\%$ Secure).  To ensure that researchers obtain access to the
revised data, we have established a
website\footnote{\url{http://www2.keck.hawaii.edu/science/tksurvey}}
to disseminate all data products resulting from this survey.  In
addition to our most recent catalog of DEIMOS photometric and
spectroscopic measurements of all targets in the GOODS-N field,
astronomers can currently obtain the wavelength-calibrated 1-D spectra
and will soon be able to download the sky-subtracted 2-D images from
which the 1-D spectra were extracted.  These resources will allow
others to measure velocity widths and spectral indices for the sample,
promoting the study of fundamental galaxy properties at moderate
redshift.

\acknowledgments 

This project was the brainchild of David Koo, who conceived of the
survey, formulated a detailed plan for executing it, and (with support
from Sandra Faber) served as its principal advocate.  The survey would
not have been completed without the generosity of the DEEP2 Redshift
Survey Team, which not only developed the sophisticated data reduction
pipeline for DEIMOS, but also donated a significant amount of their
DEIMOS observing time to complete the survey.  We are indebted to
Garth Illingworth and Pieter van Dokkum for acquiring several of the
DEIMOS images required for the photometric survey.  Additionally, we
thank Drew Phillips for helping process the slitmask designs and for
use of his cosmic ray rejection algorithm, Ben Weiner for useful
discussions and for running his redshift measuring code independently
on this sample, plus Tommaso Treu and Richard Ellis for sharing
redshifts from their GOODS-N survey prior to publication.

The National Science Foundation supported our survey through grant
AS-0331730.  This research has made use of the NASA/IPAC Extragalactic
Database (NED) which is operated by the Jet Propulsion Laboratory,
California Institute of Technology, under contract with the National
Aeronautics and Space Administration.  This research also relied upon
data from the Sloan Digital Sky Survey (SDSS).  Funding for the
creation and distribution of the SDSS Archive has been provided by the
Alfred P. Sloan Foundation, the Participating Institutions, the
National Aeronautics and Space Administration, the National Science
Foundation, the U.S. Department of Energy, the Japanese
Monbukagakusho, and the Max Planck Society.  The SDSS is managed by
the Astrophysical Research Consortium (ARC) for the Participating
Institutions. The Participating Institutions are The University of
Chicago, Fermilab, the Institute for Advanced Study, the Japan
Participation Group, The Johns Hopkins University, Los Alamos National
Laboratory, the Max-Planck-Institute for Astronomy (MPIA), the
Max-Planck-Institute for Astrophysics (MPA), New Mexico State
University, University of Pittsburgh, Princeton University, the United
States Naval Observatory, and the University of Washington.  Our
special thanks to Andy Connolly, who helped arrange our access to the
SDSS data prior to the DR1 release.

We thank the anonymous reviewer for helpful suggestions which improved
the quality of the manuscript.  C. N. A. W. thanks the W. M. Keck
Observatory for its hospitality throughout the visit during which much
of this work was completed.

We also wish to recognize and acknowledge the highly significant
cultural role and reverence that the summit of Mauna Kea has always
had within the indigenous Hawaiian community.  It is a privilege to be
given the opportunity to conduct observations from this mountain.

\begin{turnpage}
\begin{deluxetable}{rrrrrrrrrrrrrrrrrrr}
  \tabletypesize{\scriptsize}
  \tablewidth{0pt}
  \tablecaption{\label{tab-tkrs}TKRS Catalog}
  \tablehead{
    \colhead{No.\tablenotemark{a}} &
    \colhead{$\alpha$\tablenotemark{b}}               &
    \colhead{$\delta$\tablenotemark{c}}               &
    \colhead{$R_{AB}$\tablenotemark{d}}&
    \colhead{Mask\tablenotemark{e}}&
    \colhead{Slit\tablenotemark{f}}&
    \colhead{$z$\tablenotemark{g}}&
    \colhead{$Q$\tablenotemark{h}}&
    \colhead{$z_o$\tablenotemark{i}}&
    \colhead{Ref.\tablenotemark{j}}&
    \colhead{$X_d$\tablenotemark{k}}&
    \colhead{$Y_d$\tablenotemark{l}}&
    \colhead{ACS\tablenotemark{m}}&
    \colhead{$X_a$\tablenotemark{n}}&
    \colhead{$Y_a$\tablenotemark{o}}&
    \colhead{$a$\tablenotemark{p}}&
    \colhead{$e2$\tablenotemark{q}}&
    \colhead{$\Theta$\tablenotemark{r}}&
    \colhead{ACS No.\tablenotemark{s}}\\
    &\multicolumn{1}{c}{(J2000)}
    &\multicolumn{1}{c}{(J2000)}
    &&&&&&&
    &\multicolumn{1}{c}{(px)}
    &\multicolumn{1}{c}{(px)}
    &
    &\multicolumn{1}{c}{(px)}
    &\multicolumn{1}{c}{(px)}
    &\multicolumn{1}{c}{($\arcsec$)}
    &
    &\multicolumn{1}{c}{($\degr$)}}
  \startdata
1468 & 12~36~18.450 & 62~16~01.53 & 23.18 & 01 & 046 & 0.79803 & 4 & 0.7970 & 20 & 3291.7 & 466.3 & 43 & 4574.6 & 7529.9 & 0.521 & 0.008 & 63.9 &J123618.46+621601.9\\
1473 & 12~36~01.805 & 62~14~05.19 & 24.20 & 01 & 025 & 0.43592 & 4 & \nodata & \nodata & 1897.3 & 462.3 & 53 & 272.6 & 3663.6 & 0.674 & 0.225 & 133.4 &J123601.80+621405.6\\
1475 & 12~36~05.932 & 62~14~35.81 & 22.16 & 01 & 029 & 0.40836 & 4 & \nodata & \nodata & 2253.8 & 452.9 & 43 & 7490.8 & 4678.5 & 0.636 & 0.028 & 19.2 &J123605.97+621436.1\\
1479 & 12~36~25.632 & 62~16~47.94 & 24.61 & \nodata & \nodata & \nodata & \nodata & \nodata & \nodata & 3870.3 & 489.7 & 44 & 2897.1 & 886.6 & 0.372 & 0.003 & -34.8 &J123625.65+621648.3\\
1482 & 12~37~00.626 & 62~20~54.44 & 23.79 & 02 & 096 & 0.79390 & 4 &
\nodata & \nodata & 6809.1 & 473.9 & 35 & 2969.6 & 904.3 & 0.572 &
0.009 & 117.4 &J123700.59+622054.9\\
\enddata
\tablecomments{Table~\ref{tab-tkrs} is published in its entirety in
  the electronic edition of The Astronomical Journal.  A portion is
  shown here for guidance regarding its form and content.}
\tablerefs{
  (1)~\protect\cite{cow04};
  (2)~\protect\cite{coh96};
  (3)~\protect\cite{coh00};
  (8)~\protect\cite{coh01};
  (9)~\protect\cite{phi97};
  (11)~\protect\cite{daw01};
  (12)~\protect\cite{ste03};
  (15)~\protect\cite{dic98};
  (17)~\protect\cite{zep97};
  (18)~\protect\cite{cow96};
  (19)~\protect\cite{bun98}.}
\tablenotetext{a}{Serial number in TKRS catalog.}
\tablenotetext{b}{Right Ascension, derived from DEIMOS mosaic image.}
\tablenotetext{c}{Declination, derived from DEIMOS mosaic image.}
\tablenotetext{d}{$R$ magnitude in AB system, derived from DEIMOS mosaic image.}  
\tablenotetext{e}{TKRS mask number with which object was observed.}
\tablenotetext{f}{Slit number for object on TKRS mask.}
\tablenotetext{g}{Topocentric redshift measured by TKRS.}
\tablenotetext{h}{Redshift code assigned by TKRS
  (star= -1; 90\% confidence = 3; 99\% confidence= 4; unknown=1,2).} 
\tablenotetext{i}{Alternate redshift from literature.}
\tablenotetext{j}{Source of alternate redshift.}
\tablenotetext{k}{X position on the DEIMOS mosaic image.}
\tablenotetext{l}{Y position on the DEIMOS mosaic image.}
\tablenotetext{m}{Sector number in which the object lies on the HST/ACS
  mosaic image.}
\tablenotetext{n}{X position on the HST/ACS mosaic image.}
\tablenotetext{o}{Y position on the HST/ACS mosaic image.}
\tablenotetext{p}{Semi-major axis length measured by SExtractor on
  DEIMOS mosaic image.}
\tablenotetext{q}{Ellipticity measured by SExtractor on
  DEIMOS mosaic image.}
\tablenotetext{r}{Major axis position angle measured by SExtractor on 
  DEIMOS mosaic image (North=0, East=+90).}
\tablenotetext{s}{GOODS-N v1.0 catalog IAU identification of objects
  matching the DEIMOS catalogue from Giavalisco et al. (2004).}
\end{deluxetable}
\end{turnpage}
\end{document}